\documentclass[utf8]{ArxivHarvard}

\usepackage{url,hyperref,lineno,microtype,subcaption}
\usepackage[onehalfspacing]{setspace}

\usepackage{amsmath,amssymb}
\usepackage{graphicx}
\usepackage{booktabs}
\usepackage{array}
\usepackage{float}
\usepackage{placeins}
\usepackage{longtable}
\usepackage{caption}
\usepackage{pdflscape}
\usepackage{tabularray}
\usepackage{hyperref}

\makeatletter
\newcommand{\hyperlinkcite}[2]{%
  \hyperlink{cite.#2}{#1}%
}
\makeatother

\def\keyFont{\fontsize{8}{11}\helveticabold }
\def\firstAuthorLast{Perez {et~al.}}
\def\Authors{Antonio Perez$^{1,*}$, Avinash Singh$^{1}$, Jonathan Mitchell$^{2}$, Philip Swadling$^{2}$}
\def\Address{$^{1}$School of Computer Science, Faculty of Engineering and Information Technology,  University of Technology Sydney, Ultimo, NSW, Australia\\
$^{2}$Thales, Rydalmere, NSW, Australia}
\def\corrAuthor{Antonio Perez}
\def\corrEmail{antonio.perez@student.uts.edu.au; antonioperezz2091@gmail.com}

\begin{document}
\onecolumn
\firstpage{1}

\title[Human factors in MR helicopter pilot training]{Strategies to manage human factors in mixed reality helicopter pilot training: a systematic literature review} 

\author[\firstAuthorLast ]{\Authors}
\address{}
\correspondance{}

\extraAuth{}

\maketitle

\begin{abstract}
\textbf{Introduction:} Mixed Reality (MR) head-mounted displays (HMDs) may offer a cost-efficient, immersive alternative to conventional flight simulation displays, but adverse human factors, particularly cybersickness, visual fatigue, and ergonomic strain, can impair pilot performance and training effectiveness in safety-critical aviation contexts.\\ 
\textbf{Methods:} Using a systematic PRISMA-based review, we identified and synthesized evidence on human factors associated with MR/virtual reality (VR) HMD use in pilot training and analogous safety-critical simulation across 80 included sources. Human-factor drivers and mitigation strategies were synthesized into a dual-taxonomy, with mitigation approaches categorized into hardware, software, ergonomic, physiological, and psychological groups. Strategy viability was interpreted through a regulatory lens informed by aviation authority expectations.\\ 
\textbf{Results:} Cybersickness, visual strain, musculoskeletal fatigue, and sensory conflict were the most consistently reported issues. Strategies that preserved simulator fidelity, such as high-quality HMD selection, calibration, ergonomic setup, and structured onboarding, appeared more operationally suitable than approaches that altered realism or visual continuity, such as field-of-view restriction. Risk of bias appraisal identified recurring methodological limitations including convenience sampling, heterogeneous study designs, and limited preregistration, reducing confidence in the generalizability of some findings.\\
\textbf{Discussion:} Human factors remain a major barrier to MR HMD adoption in pilot training, while operational and regulatory constraints strongly shape which mitigation strategies are feasible and contextually suitable. Although much of the evidence base remains VR-derived, many findings appear transferable to MR due to shared HMD-mediated perceptual and ergonomic mechanisms. These findings provide guidance for developers, training providers, and regulators seeking to improve comfort and safety without undermining simulator fidelity.

\tiny
\keyFont{\section{Keywords:} mixed reality, virtual reality, pilot training, flight simulation, cybersickness, visual fatigue, ergonomics, head-mounted displays} 

\end{abstract}

\section{Introduction}
Mixed reality (MR) technologies are increasingly transforming training environments across a variety of sectors where high-fidelity, immersive simulations are critical to safety and skill development. Unlike virtual reality (VR), which places users entirely within a virtual environment, MR blends virtual and physical elements, allowing users to interact with both simultaneously. Both MR and VR offer advantages for immersive training \citep{Kini24}; however, this review is primarily concerned with MR head-mounted displays (HMDs), given their ability to integrate virtual content with real-world environments and their emerging, yet still underdefined role within aviation training standards. Aviation training, which traditionally relies on full flight simulators (FFS) with costly projection systems, could significantly benefit from MR HMDs due to their immersive capabilities and potential for cost efficiency \citep{Cross22, Pechlivanis23}. This review focuses specifically on helicopter flight training, as it is currently the only domain in which regulatory bodies have begun to define explicit conditions for the use of HMD-based simulation systems. Recent updates to certification frameworks by the European Union Aviation Safety Agency (EASA) have introduced requirements for VR HMD-based visual systems in helicopter simulation contexts \citep{EASA23}, providing an initial regulatory foundation for HMD integration. Accordingly, helicopter training provides a policy-relevant entry point for evaluating MR-based simulation while highlighting the absence of equivalent MR standards.

Although MR HMDs show significant promise to complement or even replace existing simulators, they remain underexplored within regulatory frameworks, with a need for qualification requirements and validation standards to be established. Recent high-end HMDs, like Varjo’s extended reality (XR) series \citep{Varjo}, reach near eye-limiting pixel-per-degree passthrough and integrate eye tracking for gaze-triggered autofocus, moving MR closer to FFS visual demands \citep{Cross22, Fussell24, Kimura24, Kini24}. Key performance markers such as wide field-of-view (FOV) \citep{Song21, Cross22}, low latency \citep{Song21}, and high resolution \citep{Cross22, Botha24} distinguish “advanced” from consumer-grade devices. While several findings discussed throughout this review may also be applicable to fixed-wing pilot training due to shared HMD-related perceptual, ergonomic, and simulation-design considerations, fixed-wing simulation is not the primary focus of this paper because equivalent HMD-specific regulatory guidance has not yet been established; however, the findings may help inform future fixed-wing MR HMD research.

The adoption of MR in helicopter simulation training is not without challenges, particularly concerning human factors in addition to critical fidelity characteristics. In the context of these emerging capabilities and regulatory developments, understanding the human-factor implications of MR systems is critical for their safe and effective integration into training environments. Immersive MR environments can present various physiological impacts, such as cybersickness \citep{Biwas24}, visual fatigue \citep{Song21}, and ergonomic strain \citep{Torrence22}, which can impact the effectiveness of training. Among these, it is useful to distinguish between related but distinct phenomena: cybersickness, simulator sickness, and visually induced motion sickness (VIMS). Cybersickness typically arises in immersive VR or MR environments, simulator sickness is associated with traditional fixed-base simulation systems, and VIMS occurs when motion perception is driven primarily by visual stimuli without corresponding physical movement. Despite these differences, symptoms of cybersickness are generally similar to motion sickness \citep{Mazloumi18} and are further exacerbated by latency and/or discrepancies between the visual and physical movement of the user \citep{Botha24}. Fatigue and mental workload are also significant concerns in extended training sessions, especially in high-stakes environments such as aviation, where sustained attention and cognitive engagement are essential \citep{Luong20}. These human-factor considerations are crucial for maintaining the well-being of the trainees and ensuring that the training outcomes meet the necessary safety and operational standards.

Given the relative scarcity of MR-specific empirical studies, it is necessary to draw upon adjacent evidence to understand these challenges. VR-based evidence is therefore included where appropriate to inform shared human-factor mechanisms. This is supported by prior work indicating that VR and MR HMDs rely on similar perceptual and sensorimotor processes \citep{Kirollos2023}, particularly in the integration of visual and vestibular cues, meaning that key drivers of human-factor challenges, such as sensory conflict, latency, and visual motion, are largely consistent across both modalities \citep{Botha24, Keshavarz19}. Empirical comparisons further suggest that while MR may attenuate the severity of effects such as cybersickness due to the presence of real-world reference cues \citep{Oh22}, the overall symptom profiles and underlying causes remain comparable. Beyond cybersickness, this transferability extends to broader human-factor domains, including simulation content factors and ergonomic factors \citep{Souchet23a}, which are largely driven by shared content design and hardware characteristics across VR and MR systems. Accordingly, this review not only synthesizes MR-specific evidence where available, but also contributes to bridging this gap by systematically interpreting VR-derived findings within an MR training context, thereby providing a more comprehensive understanding of human-factor challenges and their mitigation in emerging HMD-based aviation simulation environments.

To understand the feasibility of MR in helicopter training environments, it is necessary to consider not only technical capabilities and human-factor constraints, but also the regulatory conditions that govern simulator certification and use. Traditional pilot training simulators rely on expansive projection domes and six-axis motion bases to satisfy FFS-Level D criteria \citep{EASA12}, which drives substantial capital costs \citep{Oberhauser18, Moesl23}. In practice, budgets may compromise immersion to improve affordability. MR could reverse this trade-off by supporting full-cockpit eye-point accuracy, broad FOV, and crew-coordination cues at a fraction of the cost of dome-based systems \citep{Cross22, Pechlivanis23, Kimura24}. At the same time, any MR-based approach must remain credible as a training device, as mismatches between simulated and real-world cues can lead to negative training \citep{EASA12}. In this context, regulatory standards play a critical role in defining the conditions under which such systems can be adopted in practice. 

The EASA has established comprehensive standards for Flight Simulation Training Devices (FSTDs), including recent provisions that explicitly address the use of VR HMD-based display systems in helicopter training environments \citep{EASA23}. These standards specify requirements for cockpit representation and visual systems to preserve cross-cockpit training cues. While these provisions are defined for VR HMD-based systems, many of the underlying requirements, such as FOV, latency, and tracking stability, are equally applicable to MR HMDs, given the shared hardware architectures and perceptual constraints that govern HMD technologies. For example, recent updates require a “continuous” cross-cockpit FOV of $\pm40^\circ$ horizontally and $30^\circ$ up and $35^\circ$ down vertically, alongside constraints on head-tracking latency, alignment, and color calibration \citep{EASA23}. EASA standards also recommend motion platforms to replicate aircraft accelerations \citep{EASA12}, reducing visual–vestibular conflicts that contribute to cybersickness \citep{Kim20a}. However, the high acquisition cost and space requirements of motion platforms may limit their feasibility for widespread adoption, particularly for smaller training providers, prompting interest in lower-cost mitigation pathways that can reduce cybersickness without the associated infrastructure burden. Furthermore, requirements for stable head tracking on moving bases \citep{EASA23} may need to be reconsidered as MR hardware and tracking pipelines evolve. 

Taken together, the emergence of regulatory provisions for HMD-based helicopter simulation, combined with unresolved human-factor challenges, motivates a focused synthesis of these challenges and their mitigation strategies, alongside an evaluation of their feasibility within real-world training settings. While prior reviews have examined HMD technologies in broader flight training and simulation contexts \citep{Ross23,Somerville25}, these reviews primarily focus on training effectiveness, adoption, and educational outcomes rather than the operational suitability of mitigation strategies within emerging regulatory frameworks. Furthermore, comparatively limited attention has been given to how human-factor mitigation approaches interact with simulator fidelity requirements, procedural realism, and emerging certification constraints that influence real-world deployment feasibility. Accordingly, this review addresses the current lack of synthesized human-factor evidence specific to MR helicopter pilot training by evaluating mitigation strategies through a combined human-factor, operational, and regulatory lens.

Accordingly, this systematic review addresses the following research questions (RQs):
\begin{enumerate}
    \item What are the human-factor challenges in MR helicopter pilot training?
    \item What causes human-factor challenges in MR helicopter pilot training?
    \item Which strategies have been proposed to mitigate the identified human-factor challenges in MR helicopter pilot training?
    \item How do these strategies compare in feasibility, context suitability, and priority for real-world integration in MR helicopter pilot training?
\end{enumerate}

By synthesizing the human-factor impacts of MR HMDs in helicopter pilot training, while explicitly accounting for the limitations of VR-derived evidence, this work provides a structured understanding of the associated challenges, mitigation strategies, and practical constraints, supporting the development of improved training frameworks and future regulatory standards for MR in aviation.

\section{Methods}
\subsection{Study design and protocol}
This systematic literature review was conducted using the Preferred Reporting Items for Systematic Reviews and Meta-Analyses (PRISMA) framework \citep{Moher2009} to ensure a rigorous and comprehensive exploration of the literature. The RQs were framed using PICOS (Population, Intervention, Comparator, Outcomes, Study Designs), with the PICOS elements operationalized through the search strings, eligibility criteria and data extraction fields (Tables \ref{tab:search-strings}–\ref{tab:data-extraction}).

\subsection{Data sources and search strategy}
Searches were conducted in the Association for Computing Machinery (ACM) Digital Library, Scopus, Institute of Electrical and Electronics Engineers (IEEE) Xplore, Google Scholar, and the University of Technology Sydney (UTS) Library catalog. The search strategy employed a structured set of search strings designed to capture literature relating to VR/MR HMDs, flight simulation, human factors, and mitigation strategies. Search strings targeted key themes relevant to the objectives of this review, including cybersickness, ergonomics, fatigue, simulation fidelity, and aviation training, and were adapted to the syntax of each database. To improve specificity and reduce retrieval of non-relevant records, several broad search strings were refined during revision while preserving coverage of adjacent VR/MR human-factor literature. Broad searches prioritized sensitivity over specificity to capture transferable evidence across immersive HMD domains. Due to the broader indexing behavior of Google Scholar, retrieval counts were substantially higher than those of curated academic databases; accordingly, only the first several hundred relevance-ranked results were screened, and Google Scholar was used primarily to identify potentially relevant gray literature and adjacent studies not captured elsewhere. Retrieved records were subsequently narrowed through relevance filtering, duplicate removal, title/abstract screening, and predefined inclusion/exclusion criteria. Table \ref{tab:search-strings} lists the search strings used in this review.

\normalsize

\footnotesize

\setcounter{table}{0}
\refstepcounter{table}\label{tab:search-strings}
\edef\SearchTableNo{\thetable}

\setlength\LTleft{0pt}
\setlength\LTright{0pt}

\setlength{\tabcolsep}{2.5pt}

\begin{longtable}{p{0.57\linewidth}ccccccc}

\multicolumn{8}{@{}l@{}}{\normalsize\textbf{Table \SearchTableNo.} \textit{Search strings and database retrieval counts.}}\\[-0.25\baselineskip]
\hline
\textbf{Search String} &
\begin{tabular}[c]{@{}c@{}}\textbf{Google}\\ \textbf{Scholar$^\dagger$}\end{tabular} &
\textbf{ACM} &
\textbf{Scopus} &
\textbf{IEEE} &
\textbf{UTS} &
\textbf{Total} &
\begin{tabular}[c]{@{}c@{}}\textbf{Total}\\ \textbf{Screened}\end{tabular} \\
\hline
\endhead

\hline
\endfoot

\hline
\multicolumn{8}{p{\linewidth}}{\footnotesize $^\dagger$Due to the broader indexing behaviour of Google Scholar, only the first 300 relevance-ranked results per search string were screened.} \\
\endlastfoot

``Human Factors'' AND (``VR'' OR ``Virtual Reality'') AND (``MR'' OR ``Mixed Reality'') AND ``HMD'' 
& 16,400 & 778 & 6 & 27 & 25 & 17,236 & 1,136 \\

(``Cybersickness'' OR ``simulator sickness'' OR ``VIMS'' OR ``visually induced motion sickness'' OR ``motion sickness'') AND (``VR'' OR ``MR'') AND (``aviation'' OR ``flight simulation'' OR ``pilot training'') 
& 12,200 & 288 & 37 & 19 & 71 & 12,615 & 715 \\

``Peripheral vision'' AND ``optical flow'' AND ``immersion'' AND (``VR'' OR ``MR'') 
& 408 & 86 & 0 & 2 & 1 & 497 & 389 \\

``Psychovisual impacts'' AND (``VR'' OR ``MR'') 
& 3 & 0 & 0 & 0 & 0 & 3 & 3 \\

``User embodiment'' AND (``VR'' OR ``MR'') 
& 851 & 112 & 23 & 351 & 19 & 1,356 & 805 \\

``User fatigue'' AND ``ergonomics'' AND (``VR'' OR ``MR'') 
& 773 & 82 & 4 & 15 & 4 & 878 & 405 \\

``Vergence Accommodation Conflict'' AND (``VR'' OR ``MR'') 
& 2,340 & 169 & 109 & 19 & 109 & 2,746 & 716 \\

``Safety critical industries'' AND (``VR'' OR ``MR'') 
& 1,250 & 9 & 3 & 26 & 5 & 1,293 & 343 \\

(``VR'' OR ``Virtual Reality'') AND (``Mixed Reality'' OR ``MR'') AND ``flight simulator'' 
& 5,070 & 161 & 17 & 45 & 34 & 5,327 & 557 \\

(``VR'' OR ``Virtual Reality'') AND (``Mixed Reality'' OR ``MR'') AND ``aviation'' 
& 36,600 & 573 & 41 & 46 & 132 & 37,392 & 1,092 \\

(``VR'' OR ``Virtual Reality'') AND (``Mixed Reality'' OR ``MR'') AND ``flight training'' 
& 2,430 & 42 & 9 & 30 & 33 & 2,544 & 414 \\

(``Medication'' OR ``Pharmaceutical'') AND ``cybersickness'' AND (``VR'' OR ``Virtual Reality'') AND (``Mixed Reality'' OR ``MR'') 
& 955 & 130 & 1 & 2 & 1 & 1,089 & 434 \\

(``mitigation'' OR ``countermeasure*'' OR ``intervention*'') AND (``cybersickness'' OR ``simulator sickness'' OR ``visual fatigue'') AND (``VR'' OR ``MR'') AND (``HMD'' OR ``flight simulation'' OR ``pilot training'') 
& 6,060 & 382 & 26 & 11 & 29 & 6,508 & 748 \\

(``pilot'' OR ``pilot training'' OR ``aircrew'' OR ``flight crew'' OR ``trainee'') AND (``VR'' OR ``MR'') AND (``flight simulator'' OR ``flight simulation'' OR ``flight training'') 
& 17,500 & 360 & 135 & 67 & 326 & 18,388 & 1,188 \\

``Helicopter'' AND (``VR'' OR ``MR'') AND (``flight simulator'' OR ``flight simulation'' OR ``flight training'') 
& 9,660 & 52 & 22 & 7 & 40 & 9,781 & 421 \\

(``latency'' OR ``tracking'' OR ``field of view'' OR ``FOV'' OR ``resolution'') AND (``VR'' OR ``MR'') AND (``flight simulator'' OR ``flight simulation'') 
& 16,900 & 395 & 47 & 81 & 51 & 17,474 & 874 \\

\end{longtable}
\normalsize

\subsection{Eligibility criteria}
Peer-reviewed English articles were retained when they: (i) examined MR/VR HMDs in flight or analogous safety-critical training; (ii) reported empirical data or formal analyses on human factors such as cybersickness, fatigue, ergonomics, vergence-accommodation conflict (VAC) or peripheral-vision effects. Papers lacking human-factor content, non-peer-reviewed reports and non-English texts were excluded. An exception was made for a small number of non-peer-reviewed articles when they were deemed extremely relevant and from context-specific journals, e.g. Journal of Aviation/Aerospace Education and Research. These criteria are summarized in Tables~\ref{tab:inclusion-criteria} and~\ref{tab:exclusion-criteria}, respectively.

\scriptsize

\setcounter{table}{1}
\refstepcounter{table}\label{tab:inclusion-criteria}
\edef\InclusionTableNo{\thetable}

\setlength\LTleft{0pt}
\setlength\LTright{0pt}
\begin{longtable}{p{0.10\linewidth} p{0.85\linewidth}}

\multicolumn{2}{@{}l@{}}{\normalsize\textbf{Table \InclusionTableNo.} \textit{Inclusion criteria.}}\\[-0.25\baselineskip]
\hline
\textbf{ID} & \textbf{Inclusion Criteria (IC)} \\
\hline
\endfirsthead

\multicolumn{2}{@{}l@{}}{\normalsize\textbf{Table \InclusionTableNo\ (continued).} \textit{Inclusion criteria.}}\\[-0.25\baselineskip]
\hline
\textbf{ID} & \textbf{Inclusion Criteria (IC)} \\
\hline
\endhead

\hline
\endfoot
\hline
\endlastfoot

IC1 & Examined MR and VR HMDs in flight simulation or similar high-stakes training environments, including aviation and closely analogous safety-critical training where findings are transferable. \\
IC2 & Addressed human factors such as cybersickness, fatigue, or ergonomics in the context of MR/VR including discussion of contributing factors where available. \\
IC3 & Provided empirical data or analyses on MR/VR HMD performance and/or human-factor impacts in safety-critical or immersive simulation settings. \\
IC4 & Peer-reviewed journal or conference papers. \\

\end{longtable}
\normalsize

\scriptsize

\setcounter{table}{2}
\refstepcounter{table}\label{tab:exclusion-criteria}
\edef\ExclusionTableNo{\thetable}

\setlength\LTleft{0pt}
\setlength\LTright{0pt}
\begin{longtable}{p{0.10\linewidth} p{0.85\linewidth}}

\multicolumn{2}{@{}l@{}}{\normalsize\textbf{Table \ExclusionTableNo.} \textit{Exclusion criteria.}}\\[-0.25\baselineskip]
\hline
\textbf{ID} & \textbf{Exclusion Criteria (EC)} \\
\hline
\endfirsthead

\multicolumn{2}{@{}l@{}}{\normalsize\textbf{Table \ExclusionTableNo\ (continued).} \textit{Exclusion criteria.}}\\[-0.25\baselineskip]
\hline
\textbf{ID} & \textbf{Exclusion Criteria (EC)} \\
\hline
\endhead

\hline
\endfoot
\hline
\endlastfoot

EC1 & Focused solely on VR/MR applications unrelated to human factors or performance evaluation in simulation/training contexts. \\
EC2 & Were not peer-reviewed, including white papers or non-academic articles, to maintain a high standard of academic quality. \\
EC3 & The study was not written in English. \\

\end{longtable}
\normalsize

\subsection{Study selection}
A total of 10,240 records were identified through database searching. Due to the broader indexing behavior of Google Scholar, only the first 300 relevance-ranked results per search string were screened and included within the PRISMA identification counts. An initial screening of titles and abstracts was conducted to remove clearly irrelevant records, resulting in 124 records being retained for further consideration. Following this, four duplicate records were identified and removed, yielding 120 unique records (Figure~\ref{fig:prisma}). Four authors independently screened these records, excluding 50 on relevance, and assessed the remaining 70 full texts against eight predefined quality criteria (QC) summarized in Table~\ref{tab:quality-criteria}. The criteria were designed to capture complementary dimensions of relevance to the review objectives, including human-factor mechanisms, mitigation strategies, operational considerations, and transferability between VR and MR contexts. Studies were not required to satisfy all eight criteria for inclusion; rather, papers demonstrating a strong subset of relevant criteria were retained where they provided meaningful contributions to the review aims. For example, QC5 was included because comparisons between VR and MR are particularly relevant to this review’s regulatory framing, as current HMD-oriented aviation standards are primarily VR-based while many underlying human-factor mechanisms are transferable to MR systems. Disagreements were resolved by consensus, resulting in 48 primary studies being retained from database searches. 
To ensure comprehensive coverage of the literature, backward snowballing was performed by examining the reference lists of the 48 included studies, with a particular focus on identifying the original empirical studies cited within review articles. This process yielded an additional 32 articles. Importantly, all snowballed studies were subjected to the same inclusion and exclusion criteria, as well as the same quality assessment procedures, as the database-derived records. This approach ensured consistent study selection while enabling inclusion of primary empirical evidence not captured through database searches.
{ 
\refstepcounter{figure}\label{fig:prisma}
\edef\PrismaFigNo{\thefigure}

\begin{figure}[!htbp]
  \centering
  \includegraphics[width=\linewidth,height=0.4\textheight,keepaspectratio, alt={PRISMA flow diagram shows the literature selection process for a review: ten thousand two hundred forty records identified, one hundred twenty-four retained, four duplicates removed, fifty and twenty-two excluded, leaving eighty studies included, of which thirty-two were identified through backward snowballing of reference lists.}]{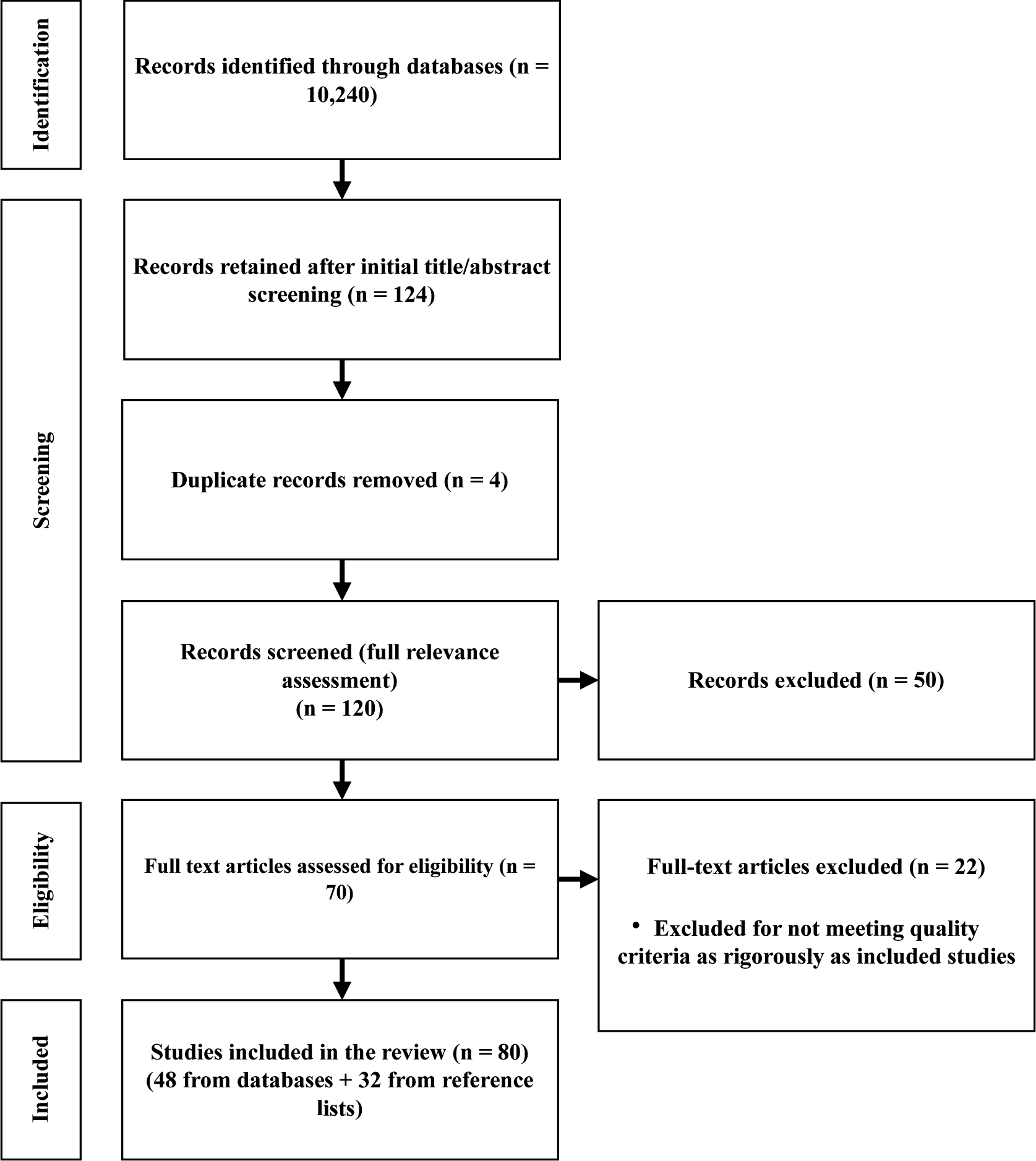}

  \vspace{0.25\baselineskip}
  \makebox[\linewidth][l]{\textbf{Figure \PrismaFigNo.} \textit{PRISMA flow diagram of our study selection process.}}
\end{figure}
} 
\FloatBarrier
\scriptsize

\setcounter{table}{3}
\refstepcounter{table}\label{tab:quality-criteria}
\edef\QualityTableNo{\thetable}

\setlength\LTleft{0pt}
\setlength\LTright{0pt}
\begin{longtable}{p{0.10\linewidth} p{0.85\linewidth}}

\multicolumn{2}{@{}l@{}}{\normalsize\textbf{Table \QualityTableNo.} \textit{Quality criteria.}}\\[-0.25\baselineskip]

\hline
\textbf{ID} & \textbf{Quality Criteria (QC)} \\
\hline
\endfirsthead

\multicolumn{2}{@{}l@{}}{\normalsize\textbf{Table \QualityTableNo\ (continued).} \textit{Quality criteria.}}\\[-0.25\baselineskip]
\hline
\textbf{ID} & \textbf{Quality Criteria (QC)} \\
\hline
\endhead

\hline
\endfoot
\hline
\endlastfoot

QC1 & Does the study address key human factors such as cybersickness, fatigue, or VAC in MR/VR HMDs and/or their determinants? \\
QC2 & Does the study focus on the use of MR/VR in safety-critical industries (e.g., aviation, medical, military) with relevance to training/simulation? \\
QC3 & Does the study provide insights into the effects of peripheral vision, immersion, or optical flow on user performance in MR/VR environments or other factors linked to human-factor outcomes? \\
QC4 & Does the study propose strategies to mitigate human-factor challenges (e.g., cybersickness, discomfort) in MR/VR environments? \\
QC5 & Does the study provide comparative insights relevant to MR and VR, including their strengths, weaknesses, transferability, or suitability for aviation training where applicable? \\
QC6 & Does the study include practical design recommendations for addressing human-factor challenges in MR/VR aviation training environments or closely related training contexts? \\
QC7 & Does the study provide evidence or discuss experimental paradigms to validate human-factor strategies in MR/VR aviation training or report evaluation approaches relevant to training outcomes? \\
QC8 & Does the study highlight challenges, lessons learned, or best practices from other safety-critical industries that have adopted MR/VR for training that may inform pilot training integration? \\

\end{longtable}
\normalsize

\subsection{Data extraction}
Data was extracted according to the criteria listed in Table~\ref{tab:data-extraction}.  For each study we recorded the simulation platform and MR/VR HMD model, participant sample and demographics, training task or scenario, the human-factor metrics assessed (e.g.\ cybersickness, fatigue, VAC, ergonomics), any mitigation strategies tested, and comparative findings versus conventional simulators or other benchmarks.  Extraction was performed by one author and independently verified by a second; all discrepancies were resolved through discussion to ensure accuracy.

\scriptsize

\setcounter{table}{4}
\refstepcounter{table}\label{tab:data-extraction}
\edef\DataExtractTableNo{\thetable}

\setlength\LTleft{0pt}
\setlength\LTright{0pt}
\begin{longtable}{p{0.10\linewidth} p{0.85\linewidth}}

\multicolumn{2}{@{}l@{}}{\normalsize\textbf{Table \DataExtractTableNo.} \textit{Data extraction criteria.}}\\[-0.25\baselineskip]
\hline
\textbf{ID} & \textbf{Data Extraction (DE)} \\
\hline
\endfirsthead

\multicolumn{2}{@{}l@{}}{\normalsize\textbf{Table \DataExtractTableNo\ (continued).} \textit{Data extraction criteria.}}\\[-0.25\baselineskip]
\hline
\textbf{ID} & \textbf{Data Extraction (DE)} \\
\hline
\endhead

\hline
\endfoot

\hline
\multicolumn{2}{@{}p{0.95\linewidth}@{}}{\footnotesize Feasibility, context suitability, and implementation priority ratings were author-derived using information reported in each study (see Section~\ref{sec:taxonomy}).} \\
\endlastfoot

DE1  & Does the paper discuss key human factors (e.g., cybersickness, fatigue, VAC) in MR/VR aviation training environments and any contributing factors where reported? \\
DE2  & Does the study propose or assess strategies to mitigate human-factor challenges in MR/VR training? \\
DE3  & What is the primary focus of the paper (e.g., MR/VR design, user performance, safety, or human factors assessment) and which RQ(s) it informs (RQ1–RQ3)? \\
DE4  & Was the methodology empirically tested or validated and what study design was used? \\
DE5  & Does the study compare MR/VR training environments with traditional flight simulators or other safety-critical industry practices or report an implicit baseline/alternative configuration? \\
DE6  & Does the study highlight ergonomic concerns (e.g., posture, HMD weight, visual fatigue) and how they affect performance and/or user experience? \\
DE7  & Does the study provide practical design recommendations for MR/VR environments to improve usability or mitigate human factors and specify the targeted issue(s)? \\
DE8  & Does the paper propose experimental paradigms or validation methods for MR/VR training strategies and specify outcome measures where available? \\
DE9  & Does the study provide lessons learned from MR/VR adoption in safety-critical industries (e.g., medical, military) that are transferable to aviation training? \\
DE10 & Does the paper address or compare different MR/VR HMDs and their impact on training effectiveness or user comfort including key system characteristics where reported? \\
DE11 & What information is reported that informed the authors’ appraisal of feasibility, context suitability, or implementation priority? \\

\end{longtable}


\normalsize

\subsection{Data analysis}
Findings were grouped into two analytic categories: (i) performance metrics (FOV, latency, resolution, tracking accuracy); (ii) human factors (cybersickness, fatigue, mental workload, peripheral vision). Narrative synthesis identified recurrent challenges and design strategies; quantitative pooling was not feasible owing to metric heterogeneity.

\subsection{Taxonomy}
\label{sec:taxonomy}
This section provides the conceptual scaffolding used to interpret the evidence that follows. Two complementary taxonomies address our RQs, with the first taxonomy (Fig.~\ref{fig:taxonomy}) primarily addressing RQ2 by organizing the etiological drivers underlying the human-factor challenges identified in RQ1, and the second taxonomy (Table~\ref{tab:strategies-template}) addressing RQ3/4. Each taxonomy is intentionally parsimonious while remaining sufficiently granular to guide design and evaluation. While cybersickness is a prominent and well-studied outcome in immersive environments, the taxonomy is designed to capture a broader set of human-factor challenges, including perceptual, cognitive, ergonomic, and interaction-related factors that influence usability, training effectiveness, and overall system performance. The first taxonomy classifies the etiological drivers of human-factor challenges (e.g., individual, hardware, simulation-content, and ergonomic factors), whereas the second taxonomy classifies mitigation strategies according to their primary implementation domain (e.g., hardware, software, ergonomic, psychological, and physiological strategies). Broadly, individual factors map most closely to psychological and physiological mitigation strategies, while simulation-content factors primarily map to software-based interventions. Accordingly, relationships between the two taxonomies are not strictly one-to-one, and some mitigation strategies may address multiple etiological categories simultaneously. These two taxonomies, presented visually in Fig.~\ref{fig:taxonomy} and operationalized in Table~\ref{tab:strategies-template}, provide the framework used in the results to map identified human-factor drivers to their corresponding mitigation strategies.
{ 
\refstepcounter{figure}\label{fig:taxonomy}
\edef\TaxonomyFigNo{\thefigure}

\begin{figure}[H]
  \centering
  \includegraphics[width=\linewidth, alt={Flowchart titled “Causes of human factors” showing four main categories: individual factors, hardware factors, simulation content factors, and ergonomic factors, each with five to six contributing elements listed under them.}]{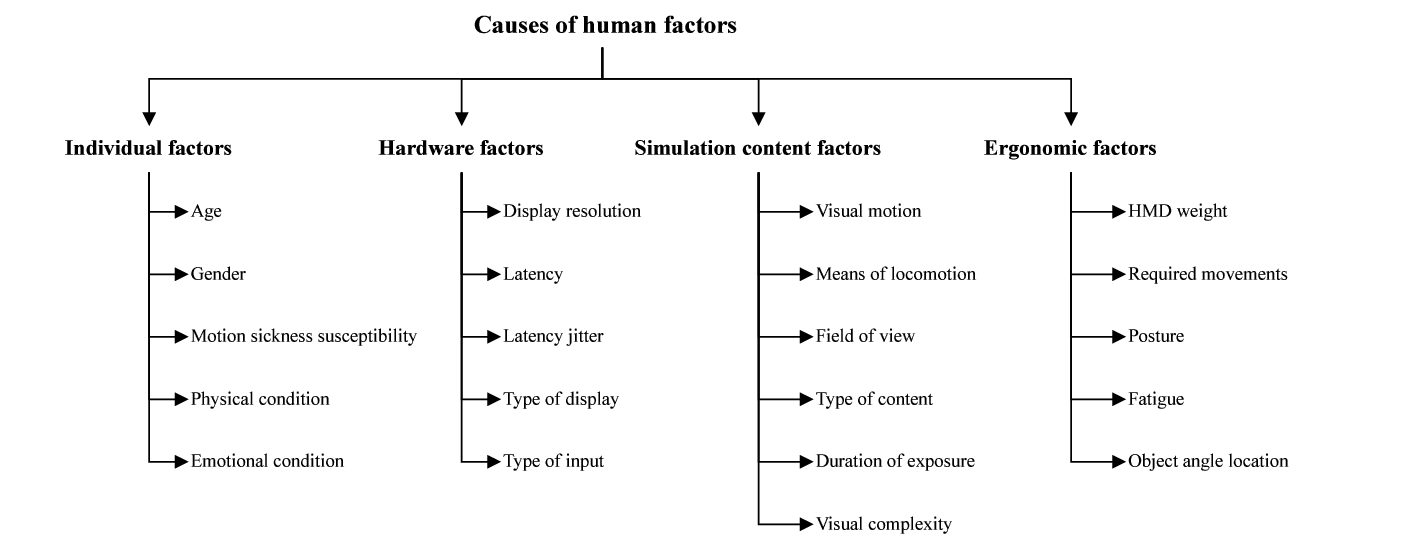}

  \vspace{0.25\baselineskip}
  \noindent\begin{minipage}{\linewidth}
  \raggedright
  \textbf{Figure \TaxonomyFigNo.} \textit{The taxonomy, modified and adapted from \citet{Biwas24}, organizes eighteen determinants into four mutually exclusive clusters.}
  \end{minipage}
\end{figure}
} 
{ 
\scriptsize

\setcounter{table}{5}
\refstepcounter{table}\label{tab:strategies-template}
\edef\MitigationTableNo{\thetable}

\setlength\LTleft{0pt}
\setlength\LTright{0pt}

\setlength{\tabcolsep}{3.5pt}
\renewcommand{\arraystretch}{1.12}

\begin{longtable}{@{}p{0.14\linewidth} p{0.44\linewidth} p{0.10\linewidth}
                  p{0.08\linewidth} p{0.10\linewidth} p{0.08\linewidth}@{}}

\multicolumn{6}{@{}l@{}}{\normalsize\textbf{Table \MitigationTableNo.} \textit{Mitigation strategies taxonomy.}}\\[-0.15\baselineskip]
\hline
\rule{0pt}{2.6ex}\textbf{Strategy} & \textbf{References} & \textbf{Category} &
\textbf{\shortstack[l]{Feasi-\\bility}} &
\textbf{\shortstack[l]{Context\\suitability}} &
\textbf{Priority} \\
\hline
\endfirsthead

\multicolumn{6}{@{}l@{}}{\normalsize\textbf{Table \MitigationTableNo\ (continued).} \textit{Mitigation strategies taxonomy.}}\\[-0.15\baselineskip]
\hline
\rule{0pt}{2.6ex}\textbf{Strategy} & \textbf{References} & \textbf{Category} &
\textbf{\shortstack[l]{Feasi-\\bility}} &
\textbf{\shortstack[l]{Context\\suitability}} &
\textbf{Priority} \\
\hline
\endhead

\hline
\endfoot

\hline
\multicolumn{6}{@{}p{\linewidth}@{}}{%
\vspace{-1em}
\footnotesize Psych = psychological; physio = physiological.
}\\
\endlastfoot

\multicolumn{3}{@{}l@{}}{%
\begin{tabular}[t]{@{}l@{}}
\textit{Categories: Hardware; Software}\\
\textit{Ergonomic; Psych; Physio}
\end{tabular}
} &
\multicolumn{3}{l@{}}{\textit{Rankings: Low; Medium; High.}}\\[24pt]

\textit{Example strategy name} &
\textit{(Author, year)} &
\textit{Software} &
\textit{Medium} &
\textit{Low} &
\textit{Low} \\

\end{longtable}

\normalsize
} 
\setcounter{table}{6}

To evaluate proposed mitigation strategies, we applied a three-axis intervention framework that considers: (1) feasibility (e.g., engineering effort, regulatory overhead), (2) context suitability (e.g., compatibility with helicopter simulation environments), and (3) implementation priority (e.g., urgency based on issue severity/frequency and ranking for feasibility/context suitability). These criteria were applied through a qualitative scoring process conducted by the authors, where each mitigation strategy was independently assessed across the three dimensions. Initial ratings were discussed collaboratively, and any discrepancies were resolved through consensus to ensure consistency in interpretation. Although the evaluation involved expert judgment, clearly defined criteria and structured discussion reduced subjectivity and improved reliability. The resulting classifications are presented in Table~\ref{tab:strategies-template} to support downstream analysis and decision-making.

\section{Results}
\label{sec:results}

\subsection{Study selection and characteristics}
The composition of the included evidence base ($N=80$) provides context for interpreting the taxonomy mappings that follow. The corpus was intentionally constructed to capture both empirical evidence of human-factor effects and practical insights into system design and implementation. Accordingly, the dataset includes empirical studies, design and technical papers, and review articles to balance measured outcomes with broader system-level considerations (Table~\ref{tab:research-design}).
This distribution indicates that the evidence base is primarily grounded in measurable outcomes such as cybersickness, task performance, and physiological responses. Furthermore, the evidence base reflects a multidisciplinary field distribution (Table~\ref{tab:field-of-research}), with human-computer interaction (HCI) and user experience (UX) venues contributing the largest share. While 15 studies are directly situated within aviation, aerospace, or flight simulation domains, a broader set of 23 studies incorporate aviation-relevant scenarios or tasks, indicating that aviation-related insights extend beyond domain-specific venues. Study-level characteristics further show that the evidence base is dominated by VR studies, with comparatively fewer MR-specific studies and a small number of VR/MR/augmented reality (AR) reviews (Table~\ref{tab:study-characteristics}). Additional contributions come from healthcare/clinical training, computer science/engineering/display optics, and psychology/behavioral science. This distribution reflects the interdisciplinary nature of the field, where technical, behavioral, and applied perspectives collectively inform the understanding of human-factor challenges. In terms of publication type, the corpus consists primarily of journal articles ($52/80$), followed by conference proceedings ($23/80$), with a smaller number of book chapters ($4/80$) and other sources ($1/80$). This is consistent with a rapidly evolving, technology-driven domain in which new methods and systems are often first reported in conference venues before being extended into journal publications. 

Overall, three implications follow for interpreting the results. First, the taxonomy mapping is well-supported for human-factor drivers and mitigation strategies that are commonly measured in controlled studies, but less so for organizational and long-term training considerations. Second, the inclusion of multiple study types enables a broader understanding of the problem space, with empirical studies being used to support causal relationships and measured effects, while design and review papers expand the range of proposed solutions and highlight areas requiring further validation. Third, the dominance of VR-based studies relative to MR-specific evidence indicates that many findings must currently be interpreted through transferable human-factor mechanisms shared across immersive HMD systems, while acknowledging that certain MR-specific characteristics and operational constraints remain underexplored.

\begin{table}[!htbp]
\setlength{\tabcolsep}{6pt}
\noindent

\begin{minipage}[t]{0.48\linewidth}
\raggedright
\refstepcounter{table}
\textbf{Table \thetable.} \textit{Primary research design of included sources ($N=80$).}
\label{tab:research-design}

\vspace{0\baselineskip}

\scriptsize
\begin{tabular}{@{}l c@{}}
  \hline
  \textbf{Research design} & \textbf{Count} \\
  \hline
  Quantitative empirical & 55 \\
  Other reviews (narrative/scoping) & 11 \\
  Design/technical/method papers & 7 \\
  Mixed methods & 3 \\
  Systematic reviews/meta-analyses & 3 \\
  Qualitative empirical & 1 \\
  \hline
  \textbf{Total} & \textbf{80} \\
  \hline
\end{tabular}
\normalsize
\end{minipage}
\hfill
\begin{minipage}[t]{0.48\linewidth}
\raggedright
\refstepcounter{table}
\textbf{Table \thetable.} \textit{Primary field of research for included sources ($N=80$).}
\label{tab:field-of-research}

\vspace{0\baselineskip}

\scriptsize
\begin{tabular}{@{}p{0.78\linewidth} c@{}}
  \hline
  \textbf{Field} & \textbf{Count} \\
  \hline
  HCI \& UX & 28 \\
  Aviation, Aerospace \& Flight Simulation & 15 \\
  Medicine, Healthcare \& Clinical Training & 10 \\
  Computer Science, Engineering \& Display Optics & 14 \\
  Psychology \& Behavioral Science & 13 \\
  \hline
  \textbf{Total} & \textbf{80} \\
  \hline
\end{tabular}
\normalsize
\end{minipage}

\end{table}

\begin{landscape}

\refstepcounter{table}\label{tab:study-characteristics}
\edef\StudyCharTableNo{\thetable}

\setlength{\tabcolsep}{3pt}
\renewcommand{\arraystretch}{1.15}

\setlength\LTleft{0pt}
\setlength\LTright{0pt}

\footnotesize
\begin{longtable}{|
p{0.12\linewidth}|
p{0.053\linewidth}|
p{0.28\linewidth}|
p{0.12\linewidth}|
p{0.053\linewidth}|
p{0.28\linewidth}|
}

\multicolumn{6}{@{}l@{}}{\normalsize\textbf{Table \StudyCharTableNo.} \textit{Study characteristics of included sources ($N=80$).}}\\[-0.25\baselineskip]
\hline

\textbf{Study} & \textbf{Immersion} & \textbf{Context} &
\textbf{Study} & \textbf{Immersion} & \textbf{Context} \\
\hline
\endfirsthead

\multicolumn{6}{@{}l@{}}{\normalsize\textbf{Table \StudyCharTableNo\ (continued).} \textit{Study characteristics of included sources ($N=80$).}}\\[-0.25\baselineskip]
\hline

\textbf{Study} & \textbf{Immersion} & \textbf{Context} &
\textbf{Study} & \textbf{Immersion} & \textbf{Context} \\
\hline
\endhead

\hline
\endfoot

\hline
\multicolumn{6}{@{}p{0.97\linewidth}@{}}{\footnotesize VR = virtual reality; MR = mixed reality; AR = augmented reality; N/A = no VR, MR, or AR HMD technology used.} \\
\endlastfoot

\cite{Alashwal21} & VR & Flying helicopter in VR under different weather conditions &
\cite{Alzayer19} & VR & FOV restriction in VR spatial navigation \\
\hline

\cite{Ang22} & VR & VR navigation study with varying terrains &
\cite{Biwas24} & VR & Systematic literature review of cybersickness in VR \\
\hline

\cite{Botha24} & VR & Framework for VR clinical simulation (no participant task) &
\cite{Chauvergne23} & VR & Study and expert interviews on onboarding methods in immersive VR systems \\
\hline

\cite{Chen21} & VR & Review of HMD ergonomics and human factors &
\cite{Cross22} & VR/AR/MR & Review of VR/MR/AR use in flight simulators \\
\hline

\cite{Curry20} & VR & VR automobile driving and third-person viewing of driving &
\cite{Damour17} & N/A & Watching first-person bicycle video, airflow and seat vibration to mitigate VIMS \\
\hline

\cite{Deluca23} & VR & Navigating human body in VR (presence/usability study) &
\cite{Fussell24} & VR & Review of VR HMD technical specifications for procedural aviation training \\
\hline

\cite{Groth21} & VR & VR racing with FOV reduction/peripheral blur &
\cite{Groth22} & VR & VR 360$^\circ$ video with galvanic vestibular stimulation (GVS) \\
\hline

\cite{Hidalgo23} & MR & MR flight simulator with cognitive monitoring &
\cite{Hildebrandt18} & VR & Redirected walking and human factors influencing cybersickness \\
\hline

\cite{Hsin23} & VR & VR medical education task &
\cite{Hussain21} & VR & Foveated depth-of-field blur in VR rollercoaster \\
\hline

\cite{Jarisch14} & N/A & Vitamin C supplement intervention to reduce seasickness &
\cite{Jasper23} & VR & Navigating VR maze \\
\hline

\cite{Kaufeld22} & VR & VR helicopter flight with chewing gum to reduce VIMS &
\cite{Keshavarz14} & N/A & Watching first-person bicycle video, pleasant music to reduce VIMS \\
\hline

\cite{Keshavarz19} & N/A & Viewing optic flow stimuli on a projector &
\cite{Kim23} & VR & Peripheral reverse optical flow in VR rollercoaster and space-flight navigation \\
\hline

\cite{Kim18} & VR & Viewing VR content with/without oculomotor exercise &
\cite{Kim20a} & VR & VR flight simulator with/without motion cues \\
\hline

\cite{Kim20b} & VR & Head movements in VR with induced display lag &
\cite{Kimura24} & MR & Electric vertical take-off and landing (eVTOL) pilot training in MR \\
\hline

\cite{Kini24} & VR/AR/MR & Review of VR/MR/AR for visual testing and gamification &
\cite{Kourtesis23} & VR & VR rides and cognitive/motor tasks \\
\hline

\cite{Krokos22} & VR & VR spaceport fly-through with electroencephalogram (EEG)-based cybersickness quantification &
\cite{Kroma21} & VR & Conditioning vs habituation in motion-sickness-inducing VR activities \\
\hline

\cite{Laudien22} & MR & MR air taxi simulator &
\cite{Lee24} & VR & Evaluation framework for cybersickness and excitement \\
\hline

\cite{Litleskare21} & VR & 360$^\circ$ VR nature walk videos and postural stability measurement &
\cite{Luong20} & VR & VR piloting task with mental workload recognition using physiological sensors \\
\hline

\cite{Mao21} & VR & VR scene exploration with nocebo effect and cybersickness framing &
\cite{Matviienko22} & VR & VR bicycle simulator \\
\hline

\cite{Gavgani17} & VR & VR rollercoaster with physiology and nausea ratings &
\cite{Mazloumi18} & VR & Cybersickness in VR rollercoaster vs classic motion sickness comparison \\
\hline

\cite{Mcanally24} & VR & Reaching/Fitts' task in VR vs real world &
\cite{Mittelstaedt18} & VR & Virtual bike simulator using HMD vs screen and ergometer/gamepad \\
\hline

\cite{Moesl23} & AR & AR-supported flight training approach/landing task &
\cite{Monteiro20} & VR & VR first person shooter (FPS) game with 2D/3D views and controllers \\
\hline

\cite{Nie19} & VR & VR racing with dynamic blurring &
\cite{Oberhauser18} & VR & VR flight simulator vs conventional simulator \\
\hline

\cite{Oh22} & VR & Varying VR content factors and cybersickness prediction &
\cite{Park17}P & VR & Oculomotor exercises before viewing VR content \\
\hline

\cite{Paroz21} & VR & VR pilot navigation game &
\cite{Pechlivanis23} & VR/MR/AR & Review of VR/MR/AR in aviation training \\
\hline

\cite{Philippe20} & VR & Review of VR learning/training applications &
\cite{Plumer23} & MR & MR drone task prototyping validation \\
\hline

\cite{Rahimi18} & VR & VR scene transitions: teleportation vs smooth locomotion &
\cite{Rahimzadeh23} & VR & Review of motion sickness mitigation in real and virtual environments \\
\hline

\cite{Rebenitsch14} & VR & VR object location &
\cite{Russell14} & VR & VR storm scene with diaphragmatic breathing to reduce motion sickness \\
\hline

\cite{Shafer17} & VR & VR FPS or space flight (piloting) game &
\cite{Shi21} & VR & VR racing game \\
\hline

\cite{Song21} & VR/AR & Review of VR and AR training applications &
\cite{Souchet23a} & VR & Review of VR ergonomic risks \\
\hline

\cite{Souchet23b} & VR & Comprehensive review of VR-induced symptoms and design guidelines &
\cite{Southgate20} & VR & Conceptual framework of embodiment in VR education \\
\hline

\cite{Sra19} & VR & VR exposure to roller coaster, car driving, leaping, and flying scenes &
\cite{Srinivasan22} & VR & Review of VR training applications in chemical safety \\
\hline

\cite{Stanney20} & VR & VR rollercoaster &
\cite{Stauffert18} & VR & VR search task with latency jitter manipulation \\
\hline

\cite{Stromberg15} & VR & VR boat simulation with controlled breathing intervention &
\cite{Thomay23} & N/A & Pilot training system using eye tracking and cognitive modeling in simulators \\
\hline

\cite{Tian22} & VR & Review of cybersickness and individual susceptibility &
\cite{Tichon14} & N/A & Flight simulation with pupillometry/electromyography (EMG) affect tracking \\
\hline

\cite{Torrence22} & VR/MR/AR & Review of VR/MR/AR applications in aviation &
\cite{Wang21} & VR/AR & VAC in VR/AR optics \\
\hline

\cite{Wang22} & VR & Predicting simulator sickness in VR games using eye and movement data &
\cite{Wang24} & MR & MR manual pointing task \\
\hline

\cite{Weech20} & VR & VR gameplay with noisy GVS &
\cite{Wienrich22} & VR & Playing VR games with virtual nose vs habituation \\
\hline

\cite{Wimmer24} & VR & VR flight simulation with EEG and pupil-based error decoding &
\cite{Won22} & VR & VR space flight video with real-time sickness detection and mitigation \\
\hline

\cite{Yadin18} & VR/AR & VAC hardware solution &
\cite{Zaghlool15} & N/A & Ginger supplement intervention to reduce nausea levels \\

\end{longtable}

\end{landscape}
\normalsize

\subsection{Risk of bias and quality appraisal across evidence types}
\label{sec:RoB}
Given the heterogeneity of the included literature, a single appraisal approach would not appropriately capture quality threats across all sources. The evidence base comprised (i) empirical studies (experimental and observational designs), (ii) design/technical/method articles, and (iii) review articles. We therefore applied three fit-for-purpose assessment approaches aligned to the main threats to validity within each evidence type. This ensured that the appraisal criteria mapped onto the methodological features that most directly affect interpretability, rather than forcing a uniform checklist across fundamentally different study genres. Empirical studies were assessed using a risk of bias (RoB) framework targeting internal and external validity (e.g., sampling, confounding, measurement quality, and reporting transparency). Design/technical/method papers were evaluated using a validation checklist focused on whether proposed systems were meaningfully tested (e.g., presence of human evaluation, comparators, and clearly defined measures). Review articles were appraised using AMSTAR 2 (``A MeaSurement Tool to Assess Systematic Reviews’’, version 2) \citep{Shea17}, which evaluates methodological rigor in evidence synthesis, including search transparency and bias assessment. AMSTAR 2 was applied consistently across review types as a conservative benchmark; lower ratings for narrative reviews reflect reduced methodological transparency rather than lack of conceptual value. 

Each assessment framework primarily used binary criterion scoring, where fulfillment of a criterion contributed one point toward the total score for that framework. The AMSTAR 2 appraisal additionally included one criterion scored on a three-level ordinal scale to better differentiate reporting quality. Total scores were then mapped to ordinal appraisal categories specific to each evidence type, where higher scores reflected lower RoB, stronger validation quality, or higher methodological confidence depending on the framework applied (Tables~\ref{tab:rob_empirical}–\ref{tab:amstar2}). The RoB and quality appraisals were performed by the first author and independently verified by the second author via a random subset in each assessment approach. Any inconsistencies were resolved by discussion and moderation of scores.

\subsubsection{Empirical articles: RoB trends}
Empirical studies (Table~\ref{tab:rob_empirical}) showed a consistent pattern in which external validity and reporting transparency were the primary limitations, while measurement and analytical rigor were generally strong. The most common weaknesses were convenience sampling without addressing selection bias, absence of preregistration, limited reporting of missing data or attrition, and inconsistent control of potential confounding variables. In contrast, most studies employed valid and reliable measures, applied them consistently, and used appropriate analytical methods. Overall, the empirical evidence base is methodologically sound at the level of measurement and analysis but constrained in generalizability and transparency, which should be considered when interpreting reported human-factor effects.

\subsubsection{Design/technical/method articles: validation quality trends}
Design/technical/method papers (Table~\ref{tab:validation_quality}) were characterized by strong system description but variable validation quality. While most studies clearly defined tasks and scenarios, comparative evaluation was often limited due to the absence of baselines or alternative conditions, increasing the risk of overestimating system effectiveness. Human evaluation was also inconsistent, reducing confidence in claims related to usability and training effectiveness. Reporting of measures, analysis, and limitations varied across studies, introducing potential reporting and validation bias. Consequently, this body of work is valuable for identifying potential solutions and system designs, but the strength of supporting evidence depends on the extent of validation and transparency in evaluation.

\subsubsection{Review articles: AMSTAR~2 quality trends}
The AMSTAR 2 appraisal (Table~\ref{tab:amstar2}) indicates that review-level evidence is primarily limited by insufficient methodological transparency and bias control. Across reviews, protocol registration and formal RoB assessment of included studies were largely absent, increasing susceptibility to selection and reporting bias and reducing confidence in how conclusions reflect underlying evidence quality. Systematic reviews demonstrated stronger methodological conduct (e.g., explicit search strategies, structured selection processes), resulting in moderate confidence ratings. In contrast, narrative and scoping reviews frequently lacked transparent reporting of search and selection procedures, leading to predominantly low or critically low confidence ratings. Accordingly, systematic reviews were treated as more reliable evidence summaries, while narrative reviews were interpreted as context-setting and hypothesis-generating rather than definitive sources of evidence.

\begin{landscape}

\setcounter{table}{9} 

\refstepcounter{table}\label{tab:rob_empirical}
\edef\RoBTableNo{\thetable}

\small
\setlength{\tabcolsep}{3pt}
\renewcommand{\arraystretch}{1.15}

\setlength\LTleft{0pt}
\setlength\LTright{0pt}

\footnotesize
\begin{longtable}{|
p{0.1\linewidth}|
p{0.03\linewidth}|
p{0.055\linewidth}|
p{0.095\linewidth}|
p{0.045\linewidth}|
p{0.045\linewidth}|
p{0.045\linewidth}|
p{0.045\linewidth}|
p{0.045\linewidth}|
p{0.045\linewidth}|
p{0.045\linewidth}|
p{0.045\linewidth}|
p{0.045\linewidth}|
p{0.045\linewidth}|
p{0.045\linewidth}|
p{0.035\linewidth}|
p{0.035\linewidth}|
}

\multicolumn{17}{@{}l@{}}{\normalsize\textbf{Table \RoBTableNo.} \textit{Empirical articles RoB assessment.}}\\[-0.25\baselineskip]
\hline

\textbf{Study} & \textbf{Type} & \textbf{Design} & \textbf{Comparator} & \textbf{Popula\-tion defined} & \textbf{Recrui\-tment appropr\-iate} & \textbf{Selection bias addressed} & \textbf{Allocation / order effects controlled} & \textbf{Confoun\-ding addressed} & \textbf{Measures valid \& reliable} & \textbf{Measure consiste\-ncy} & \textbf{Missing data addressed} & \textbf{Report\-ing bias low} & \textbf{Appropri\-ate analysis} & \textbf{Reprodu\-cibility} & \textbf{Score / 11} & \textbf{Over\-all RoB} \\
\hline
\endfirsthead

\multicolumn{17}{@{}l@{}}{\normalsize\textbf{Table \RoBTableNo\ (continued).} \textit{Empirical articles RoB assessment.}}\\[-0.25\baselineskip]
\hline
\textbf{Study} & \textbf{Type} & \textbf{Design} & \textbf{Comparator} & \textbf{Popula\-tion defined} & \textbf{Recrui\-tment appropr\-iate} & \textbf{Selection bias addressed} & \textbf{Allocation / order effects controlled} & \textbf{Confoun\-ding addressed} & \textbf{Measures valid \& reliable} & \textbf{Measure consiste\-ncy} & \textbf{Missing data addressed} & \textbf{Report\-ing bias low} & \textbf{Appropri\-ate analysis} & \textbf{Reprodu\-cibility} & \textbf{Score / 11} & \textbf{Over\-all RoB}
\\
\hline
\endhead

\hline
\endfoot

\hline
\multicolumn{17}{@{}p{0.97\linewidth}@{}}{\footnotesize Scoring convention is 'Yes' = 1 point from columns 5 to 15. Quant = quantitative; Qual = qualitative; Exp = experimental; Obs = observational; WP = within-participants; BP = between-participants; CB = counterbalanced; R = randomized; NR = non-randomized; L/M/H = low/moderate/high concern.} \\
\endlastfoot

\cite{Alashwal21} & Quant & Exp (R) & Yes — WP & Yes & Yes & No & Yes & No & Yes & Yes & No & No & Yes & Yes & 7 & M \\
\hline
\cite{Alzayer19} & Quant & Exp (NR) & Yes — mixed (WP × BP) & Yes & No & No & Yes & No & Yes & Yes & No & No & Yes & Yes & 6 & H \\
\hline
\cite{Ang22} & Quant & Exp (NR) & Yes — WP & Yes & Yes & No & Yes & Yes & Yes & Yes & Yes & No & Yes & Yes & 9 & L \\
\hline
\cite{Curry20} & Quant & Exp (NR) & Yes — BP & Yes & Yes & No & Yes & Yes & Yes & Yes & Yes & No & Yes & Yes & 9 & L \\
\hline
\cite{Damour17} & Quant & Exp (R) & Yes — BP & Yes & Yes & No & Yes & Yes & Yes & Yes & Yes & No & Yes & Yes & 9 & L \\
\hline
\cite{Deluca23} & Quant & Obs (post-hoc) & Yes — BP & No & No & No & No & No & Yes & Yes & No & No & Yes & Yes & 4 & H \\
\hline
\cite{Groth21} & Quant & Exp (R) & Yes — WP & Yes & No & No & Yes & Yes & Yes & Yes & No & No & Yes & Yes & 7 & M \\
\hline
\cite{Groth22} & Quant & Exp (NR) & Yes — WP & Yes & No & No & Yes & Yes & Yes & Yes & No & No & Yes & Yes & 7 & M \\
\hline
\cite{Hidalgo23} & Quant & Exp (NR) & No & Yes & No & No & No & No & No & Yes & No & No & No & Yes & 3 & H \\
\hline
\cite{Hildebrandt18} & Mixed & Exp (NR) & Yes — WP & Yes & No & No & No & Yes & Yes & Yes & Yes & No & Yes & No & 6 & H \\
\hline
\cite{Hsin23} & Quant & Obs(NR) & Yes — BP & Yes & Yes & No & No & Yes & Yes & Yes & No & No & Yes & Yes & 7 & M \\
\hline
\cite{Hussain21} & Quant & Exp (R) & Yes — WP & Yes & Yes & No & Yes & Yes & Yes & Yes & Yes & No & Yes & Yes & 9 & L \\
\hline
\cite{Jarisch14} & Quant & Exp (R) & Yes — BP & Yes & No & No & Yes & No & Yes & Yes & Yes & No & Yes & Yes & 7 & M \\
\hline
\cite{Jasper23} & Quant & Exp (NR) & Yes — BP & Yes & No & No & No & No & Yes & Yes & Yes & No & Yes & Yes & 6 & H \\
\hline
\cite{Kaufeld22} & Quant & Exp (R) & Yes — BP & Yes & Yes & No & Yes & Yes & Yes & Yes & Yes & Yes & Yes & Yes & 10 & L \\
\hline
\cite{Keshavarz14} & Quant & Exp (R) & Yes — BP & Yes & No & No & Yes & Yes & Yes & Yes & Yes & No & Yes & Yes & 8 & M \\
\hline
\cite{Keshavarz19} & Quant & Exp (CB) & Yes — WP & Yes & No & No & Yes & Yes & Yes & Yes & No & No & Yes & Yes & 7 & M \\
\hline
\cite{Kim23} & Quant & Exp (R) & Yes — WP & Yes & Yes & No & Yes & Yes & Yes & Yes & No & No & Yes & Yes & 8 & M \\
\hline
\cite{Kim18} & Quant & Exp (NR) & Yes — WP & Yes & No & No & No & No & Yes & Yes & No & No & Yes & Yes & 5 & H \\
\hline
\cite{Kim20a} & Quant & Exp (NR) & Yes — BP & Yes & No & No & No & No & Yes & Yes & Yes & No & Yes & Yes & 6 & H \\
\hline
\cite{Kim20b} & Quant & Exp (R) & Yes — WP & Yes & No & No & Yes & Yes & Yes & Yes & Yes & No & Yes & Yes & 8 & M \\
\hline
\cite{Kimura24} & Qual & Exp (NR) & No & Yes & Yes & No & No & No & Yes & Yes & No & No & Yes & Yes & 6 & H \\
\hline
\cite{Kourtesis23} & Quant & Exp (R) & Yes — WP & No & No & No & Yes & Yes & Yes & Yes & No & No & Yes & Yes & 6 & H \\
\hline
\cite{Krokos22} & Quant & Exp (NR) & Yes — WP & Yes & Yes & No & No & Yes & Yes & Yes & Yes & No & Yes & Yes & 8 & M \\
\hline
\cite{Kroma21} & Quant & Exp (NR) & Yes — BP & Yes & No & No & No & No & Yes & Yes & No & No & Yes & No & 4 & H \\
\hline
\cite{Laudien22} & Quant & Exp (NR) & Yes — WP & Yes & No & No & Yes & Yes & Yes & Yes & No & No & Yes & Yes & 7 & M \\
\hline
\cite{Litleskare21} & Quant & Exp (pre–post) & Yes — BP & Yes & No & No & Yes & Yes & Yes & Yes & Yes & No & Yes & Yes & 8 & M \\
\hline
\cite{Luong20} & Quant & Exp (NR) & Yes — WP & Yes & Yes & No & Yes & Yes & Yes & Yes & Yes & No & Yes & Yes & 9 & L \\
\hline
\cite{Mao21} & Quant & Exp (R) & Yes — BP & Yes & Yes & No & Yes & Yes & Yes & Yes & Yes & No & Yes & Yes & 9 & L \\
\hline
\cite{Matviienko22} & Quant & Exp (R) & Yes — WP & Yes & Yes & No & Yes & Yes & Yes & Yes & No & No & Yes & Yes & 8 & M \\
\hline
\cite{Gavgani17} & Quant & Exp (NR) & Yes — WP & Yes & No & No & No & Yes & Yes & Yes & Yes & No & Yes & Yes & 7 & M \\
\hline
\cite{Mazloumi18} & Quant & Exp (NR) & Yes — WP & Yes & No & No & No & No & Yes & Yes & No & No & Yes & Yes & 5 & H \\
\hline
\cite{Mcanally24} & Quant & Exp (CB) & Yes — WP & Yes & No & No & Yes & Yes & Yes & Yes & Yes & No & Yes & Yes & 8 & M \\
\hline
\cite{Mittelstaedt18} & Quant & Exp (R) & Yes — BP & Yes & Yes & No & Yes & Yes & Yes & Yes & No & No & Yes & Yes & 8 & M \\
\hline
\cite{Moesl23} & Quant & Exp (R) & Yes — BP & Yes & Yes & No & Yes & Yes & Yes & Yes & No & No & Yes & Yes & 8 & M \\
\hline
\cite{Monteiro20} & Quant & Exp (NR) & Yes — WP & No & No & No & Yes & No & Yes & Yes & No & No & Yes & Yes & 5 & H \\
\hline
\cite{Nie19} & Quant & Exp (R) & Yes — BP & Yes & Yes & No & Yes & Yes & Yes & Yes & Yes & No & Yes & Yes & 9 & L \\
\hline
\cite{Oberhauser18} & Quant & Exp (R) & Yes — WP & Yes & Yes & No & Yes & Yes & Yes & Yes & No & No & Yes & Yes & 8 & M \\
\hline
\cite{Oh22} & Quant & Exp (R) & Yes — WP & Yes & Yes & No & Yes & No & Yes & Yes & Yes & No & Yes & Yes & 8 & M \\
\hline
\cite{Park17} & Quant & Exp (NR) & Yes — WP & Yes & No & No & No & No & Yes & Yes & No & No & Yes & Yes & 5 & H \\
\hline
\cite{Paroz21} & Mixed & Exp (NR) & Yes — WP & Yes & Yes & No & Yes & No & Yes & Yes & No & No & Yes & Yes & 7 & M \\
\hline
\cite{Plumer23} & Quant & Exp (R) & Yes — WP & Yes & No & No & Yes & No & Yes & Yes & No & No & Yes & Yes & 6 & H \\
\hline
\cite{Rahimi18} & Quant & Exp (NR) & Yes — WP & Yes & Yes & No & Yes & Yes & Yes & Yes & No & No & Yes & Yes & 8 & M \\
\hline
\cite{Rebenitsch14} & Quant & Exp (NR) & Yes — WP & Yes & Yes & Yes  & Yes & No & Yes & Yes & No & No & Yes & Yes & 8 & M \\
\hline
\cite{Russell14} & Quant & Exp (R) & Yes — BP & Yes & Yes & No & Yes & Yes & Yes & Yes & Yes & No & Yes & Yes & 9 & L \\
\hline
\cite{Shafer17} & Quant & Exp (R) & Yes — BP & Yes & Yes & No & Yes & No & Yes & Yes & No & No & Yes & Yes & 7 & M \\
\hline
\cite{Shi21} & Quant & Exp (NR) & Yes — mixed (WP × BP) & Yes & No & No & Yes & No & Yes & Yes & No & No & Yes & Yes & 6 & H \\
\hline
\cite{Sra19} & Mixed & Exp (NR) & Yes — WP & Yes & No & No & Yes & No & Yes & Yes & No & No & Yes & Yes & 6 & H \\
\hline
\cite{Stanney20} & Quant & Exp (R) & Yes — BP & Yes & Yes & No & Yes & No & Yes & Yes & Yes & No & Yes & Yes & 8 & M \\
\hline
\cite{Stauffert18} & Quant & Exp (NR) & Yes — BP & Yes & No & No & No & No & Yes & Yes & No & No & Yes & Yes & 5 & H \\
\hline
\cite{Stromberg15} & Quant & Exp (R) & Yes — BP & Yes & Yes & No & Yes & Yes & Yes & Yes & Yes & No & Yes & Yes & 9 & L \\
\hline
\cite{Tichon14} & Quant & Exp (NR) & Yes — WP & Yes & Yes & No & No & No & Yes & Yes & No & No & Yes & Yes & 6 & H \\
\hline
\cite{Wang22} & Quant & Exp (NR) & Yes — WP & Yes & No & No & Yes & Yes & No & Yes & Yes & No & Yes & Yes & 7 & M \\
\hline
\cite{Wang24} & Quant & Exp (NR) & Yes — BP & Yes & No & No & No & No & Yes & Yes & No & No & Yes & Yes & 5 & H \\
\hline
\cite{Weech20} & Quant & Exp (R) & Yes — BP & Yes & No & No & Yes & Yes & Yes & Yes & No & No & Yes & Yes & 7 & M \\
\hline
\cite{Wienrich22} & Quant & Exp (NR) & Yes — mixed (WP \& BP) & Yes & No & No & No & Yes & Yes & Yes & No & No & No & Yes & 5 & H \\
\hline
\cite{Wimmer24} & Quant & Exp (R) & Yes — WP & Yes & No & No & Yes & Yes & Yes & Yes & Yes & No & Yes & Yes & 8 & M \\
\hline
\cite{Won22} & Quant & Exp (NR) & Yes — WP & Yes & No & No & Yes & Yes & Yes & Yes & No & No & Yes & Yes & 7 & M \\
\hline
\cite{Zaghlool15} & Quant & Exp (R) & Yes — BP & Yes & No & No & No & No & Yes & Yes & No & No & Yes & Yes & 5 & H \\

\end{longtable}

\end{landscape}
\normalsize

\begin{landscape}
\setcounter{table}{10}
\refstepcounter{table}\label{tab:validation_quality}
\edef\ValidationTableNo{\thetable}

\footnotesize
\setlength{\tabcolsep}{2pt}
\renewcommand{\arraystretch}{1.05}
\setlength{\arrayrulewidth}{0.35pt}

\setlength\LTleft{0pt}
\setlength\LTright{0pt}

\begin{longtable}{|
p{0.12\linewidth}|  
p{0.085\linewidth}| 
p{0.085\linewidth}| 
p{0.085\linewidth}|  
p{0.085\linewidth}|  
p{0.085\linewidth}|  
p{0.085\linewidth}|  
p{0.085\linewidth}|  
p{0.085\linewidth}|  
p{0.085\linewidth}|  
}

\multicolumn{10}{@{}l@{}}{\normalsize\textbf{Table \ValidationTableNo.} \textit{Design/technical/method articles validation quality assessment.}}\\[-0.25\baselineskip]
\hline

\textbf{Study} &
\textbf{Human evaluati\-on} &
\textbf{Sample described} &
\textbf{Task / scenario described} &
\textbf{Measures defined} &
\textbf{Baseline / comparator stated} &
\textbf{Analysis described} &
\textbf{Limitations / confounds discussed} &
\textbf{Score / 6} &
\textbf{Overall validation quality}
\\
\hline
\endfirsthead

\multicolumn{10}{@{}l@{}}{\normalsize\textbf{Table \ValidationTableNo\ (continued).} \textit{Design/technical/method articles validation quality assessment.}}\\[-0.25\baselineskip]
\hline
\textbf{Study} &
\textbf{Human evaluati\-on} &
\textbf{Sample described} &
\textbf{Task / scenario described} &
\textbf{Measures defined} &
\textbf{Baseline / comparator stated} &
\textbf{Analysis described} &
\textbf{Limitations / confounds discussed} &
\textbf{Score / 6} &
\textbf{Overall validation quality}
\\
\hline
\endhead

\hline
\endfoot

\hline
\multicolumn{10}{@{}p{0.97\linewidth}@{}}{\footnotesize Scoring convention is 'Yes' = 1 point from columns 3 to 8. Overall validation quality uses L/M/H = low/moderate/high. N/A indicates not applicable.}\\
\endlastfoot

\cite{Botha24} & No & N/A & Yes & Yes & N/A & Yes & No & 3 & M \\
\hline
\cite{Chauvergne23} & Yes & Yes & Yes & Yes & No & Yes & Yes & 5 & H \\
\hline
\cite{Lee24} & Yes & Yes & Yes & Yes & No & Yes & Yes & 5 & H \\
\hline
\cite{Pechlivanis23} & No & N/A & Yes & No & No & No & Yes & 2 & L \\
\hline
\cite{Southgate20} & Yes & Yes & Yes & Yes & No & Yes & No & 4 & M \\
\hline
\cite{Thomay23} & Yes & Yes & Yes & Yes & No & Yes & Yes & 5 & H \\
\hline
\cite{Yadin18} & No & N/A & Yes & Yes & Yes & Yes & No & 4 & M \\
\hline

\end{longtable}

\end{landscape}

\begin{landscape}
\setcounter{table}{11}
\refstepcounter{table}\label{tab:amstar2}
\edef\AMSTARTableNo{\thetable}

\footnotesize
\setlength{\tabcolsep}{2pt}
\renewcommand{\arraystretch}{1.05}
\setlength{\arrayrulewidth}{0.35pt}

\setlength\LTleft{0pt}
\setlength\LTright{0pt}

\begin{longtable}{|
p{0.04\linewidth}|  
p{0.04\linewidth}|  
p{0.034\linewidth}|  
p{0.034\linewidth}|  
p{0.034\linewidth}|  
p{0.034\linewidth}|  
p{0.034\linewidth}|  
p{0.034\linewidth}|  
p{0.034\linewidth}|  
p{0.034\linewidth}|  
p{0.034\linewidth}|  
p{0.034\linewidth}|  
p{0.034\linewidth}|  
p{0.034\linewidth}|  
p{0.034\linewidth}|  
p{0.034\linewidth}|  
p{0.034\linewidth}|  
p{0.034\linewidth}|  
p{0.034\linewidth}|  
p{0.034\linewidth}|  
p{0.034\linewidth}|  
p{0.034\linewidth}|  
p{0.034\linewidth}|  
p{0.034\linewidth}|  
}

\multicolumn{24}{@{}l@{}}{\normalsize\textbf{Table \AMSTARTableNo.} \textit{Review articles AMSTAR 2 quality assessment.}}\\[-0.25\baselineskip]
\hline

\textbf{Study} &
\textbf{Review type} &
\textbf{Meta-analy\-sis} &
\textbf{PICO / quest\-ion} &
\textbf{Search strate\-gy} &
\textbf{Design inclus\-ion} &
\textbf{Eligib\-ility} &
\textbf{Select\-ion} &
\textbf{Data extrac\-tion} &
\textbf{Limit\-ations} &
\textbf{Repor\-ting compl\-eteness (H = 2 / M = 1 / L = 0)} &
\textbf{Proto\-col regist\-ered} &
\textbf{Comp\-rehen\-sive search} &
\textbf{Dup. screen\-ing} &
\textbf{Dup. extrac\-tion} &
\textbf{Exclu\-ded + reasons} &
\textbf{Study charac\-teristics} &
\textbf{Fund\-ing (inclu\-ded)} &
\textbf{COI / funding} &
\textbf{RoB asses\-sed} &
\textbf{RoB in discus\-sion} &
\textbf{Heter\-ogenei\-ty} &
\textbf{Score / 20} &
\textbf{AMS\-TAR 2 confid\-ence}
\\
\hline
\endfirsthead

\multicolumn{24}{@{}l@{}}{\normalsize\textbf{Table \AMSTARTableNo\ (continued).} \textit{Review articles AMSTAR 2 quality assessment.}}\\[-0.25\baselineskip]
\hline
\textbf{Study} &
\textbf{Review type} &
\textbf{Meta-analy\-sis} &
\textbf{PICO / quest\-ion} &
\textbf{Search strate\-gy} &
\textbf{Design inclus\-ion} &
\textbf{Eligib\-ility} &
\textbf{Select\-ion} &
\textbf{Data extrac\-tion} &
\textbf{Limit\-ations} &
\textbf{Repor\-ting compl\-eteness (H = 2 / M = 1 / L = 0)} &
\textbf{Proto\-col regist\-ered} &
\textbf{Comp\-rehen\-sive search} &
\textbf{Dup. screen\-ing} &
\textbf{Dup. extrac\-tion} &
\textbf{Exclu\-ded + reasons} &
\textbf{Study charac\-teristics} &
\textbf{Fund\-ing (inclu\-ded)} &
\textbf{COI / funding} &
\textbf{RoB asses\-sed} &
\textbf{RoB in discus\-sion} &
\textbf{Heter\-ogenei\-ty} &
\textbf{Score / 20} &
\textbf{AMS\-TAR 2 confid\-ence}
\\
\hline
\endhead

\hline
\endfoot

\hline
\multicolumn{24}{@{}p{0.97\linewidth}@{}}{\footnotesize Scoring convention is 'Yes' = 1 point from columns 4 to 22 (excluding column 11). Reporting completeness is scored as high (H) = 2, medium (M) = 1, low (L) = 0. AMSTAR 2 confidence: H = high, M = moderate, L = low, CL = critically low. Dup. = duplicate; COI = conflict of interest.}\\
\endlastfoot

\cite{Biwas24} & System\-atic review & No & Yes & Yes & Yes & Yes & Yes & Yes & Yes & H & No & Yes & Yes & No & No & Yes & No & Yes & No & No & No & 13 & M \\
\hline
\cite{Chen21} & Narrative review & No & Yes & No & No & No & No & No & Yes & L & No & No & No & No & No & Yes & No & Yes & No & No & No & 4 & CL \\
\hline
\cite{Cross22} & System\-atic review & No & Yes & Yes & Yes & Yes & Yes & Yes & Yes & M & No & Yes & Yes & No & No & Yes & No & No & No & No & Yes & 12 & M \\
\hline
\cite{Fussell24} & Narrative review & No & Yes & No & No & No & No & No & No & L & No & No & No & No & No & No & No & Yes & No & No & No & 2 & CL \\
\hline
\cite{Kini24} & Narrative / scoping review & No & Yes & Yes & Yes & Yes & Yes & Yes & Yes & M & No & No & Yes & No & No & Yes & No & Yes & No & No & No & 11 & M \\
\hline
\cite{Philippe20} & Narrative review + case-study overview & No & No & Yes & No & No & No & No & Yes & L & No & No & No & No & No & No & No & Yes & No & No & No & 3 & CL \\
\hline
\hyperlinkcite{Rahimz\-adeh et al.\ (2023)}{Rahimzadeh23} & Narrative / compre\-hensive review & No & Yes & No & No & No & No & No & Yes & L & No & No & No & No & No & Yes & No & Yes & No & No & No & 4 & CL \\
\hline
\cite{Song21} & Narrative review & No & No & No & No & No & No & No & No & L & No & No & No & No & No & No & No & No & No & No & No & 0 & CL \\
\hline
\cite{Souchet23a} & Narrative review & No & Yes & Yes & Yes & Yes & Yes & Yes & Yes & M & No & No & No & No & No & No & No & Yes & No & No & Yes & 10 & CL \\
\hline
\cite{Souchet23b} & Narrative / compre\-hensive review & No & Yes & Yes & Yes & Yes & Yes & No & Yes & M & No & No & No & No & No & Yes & No & Yes & No & No & No & 9 & L \\
\hline
\cite{Srinivasan22} & Narrative review & No & Yes & No & No & No & No & No & Yes & M & No & No & No & No & No & Yes & No & Yes & No & No & No & 5 & CL \\
\hline
\cite{Tian22} & System\-atic review & No & Yes & Yes & Yes & Yes & Yes & Yes & Yes & M & No & Yes & Yes & No & No & Yes & No & Yes & No & No & Yes & 13 & M \\
\hline
\cite{Torrence22} & Narrative / critical review & No & Yes & Yes & Yes & Yes & No & Yes & Yes & M & No & No & No & No & No & Yes & No & No & No & No & No & 8 & L \\
\hline
\cite{Wang21} & Narrative review & No & Yes & No & No & No & No & No & Yes & M & No & No & No & No & No & No & No & Yes & No & No & No & 4 & CL \\
\hline

\end{longtable}

\end{landscape}
\normalsize
\subsection{Human-factor challenges and mitigation strategies in MR helicopter pilot training}
\label{sec:drivers}
The etiological taxonomy (Fig.\,\ref{fig:taxonomy}) distinguishes four mutually exclusive clusters of drivers contributing to human-factor challenges in MR HMD-based helicopter pilot training. Each factor is synthesized below alongside the mitigation strategies most relevant to addressing its associated challenges, with consideration given to feasibility, contextual suitability, operational realism, and the strength of the supporting evidence base. Across the included studies, the most consistently identified challenges related to cybersickness, sensory conflict, visual fatigue, and simulation-design factors linked to immersion, optical flow, and visual realism, reflected by the comparatively large number of software and physiological mitigation strategies identified in the literature. Hardware and ergonomic factors, including display limitations, HMD comfort, and user fatigue, were also consistently recognized despite a smaller range of mitigation approaches. Psychological and emotional influences on user adaptation and experience were represented by a comparatively smaller body of literature, although they remained a recurring consideration across immersive training contexts. Collectively, these findings address RQ1 by demonstrating that human-factor challenges in MR helicopter pilot training extend beyond cybersickness alone and emerge from interacting perceptual, physiological, psychological, ergonomic, hardware, and simulation-design mechanisms across immersive HMD systems. Particular attention is given to how findings from predominantly VR-derived literature may transfer to MR aviation contexts, as well as how EASA-aligned fidelity requirements shape the practical applicability of proposed interventions. A summary of all mitigation strategies can be found in Table~\ref{tab:strategies}.

{ 
\scriptsize

\setcounter{table}{12}
\refstepcounter{table}\label{tab:strategies}
\edef\MitigationFullTableNo{\thetable}

\setlength\LTleft{0pt}
\setlength\LTright{0pt}

\setlength{\tabcolsep}{3.5pt}
\renewcommand{\arraystretch}{1.12}

\begin{longtable}{@{}p{0.14\linewidth} p{0.44\linewidth} p{0.1\linewidth}
                  p{0.08\linewidth} p{0.10\linewidth} p{0.08\linewidth}@{}}

\multicolumn{6}{@{}l@{}}{\normalsize\textbf{Table \MitigationFullTableNo.} \textit{Mitigation strategies.}}\\[-0.15\baselineskip]
\hline
\rule{0pt}{2.6ex}\textbf{Strategy} & \textbf{References} & \textbf{Category} &
\textbf{\shortstack[l]{Feasi-\\bility}} & 
\textbf{\shortstack[l]{Context\\suitability}} &
\textbf{Priority} \\
\hline
\endfirsthead

\multicolumn{6}{@{}l@{}}{\normalsize\textbf{Table \MitigationFullTableNo\ (continued).} \textit{Mitigation strategies.}}\\[-0.15\baselineskip]
\hline
\rule{0pt}{2.6ex}\textbf{Strategy} & \textbf{References} & \textbf{Category} &
\textbf{\shortstack[l]{Feasi-\\bility}} &   
\textbf{\shortstack[l]{Context\\suitability}} &
\textbf{Priority} \\
\hline
\endhead

\hline
\endfoot
\hline
\multicolumn{6}{@{}p{\linewidth}@{}}{\footnotesize Psych = psychological, physio = physiological.} \\
\endlastfoot

Haptic Feedback & \citep{Cross22, Kimura24} & Hardware & Medium & Low & Low \\
Motion Platform & \citep{Kim20a} & Hardware & Low & High & Medium \\
HMD Choice & \citep{Yadin18, Wang21, Cross22, Souchet23b, Biwas24, Botha24} & Hardware & High & High & High \\
Input Type & \citep{Monteiro20, Laudien22, Cross22, Kimura24} & Hardware & High & High & High \\
Biometric Monitoring & \citep{Tichon14, Luong20, Philippe20, Krokos22, Oh22, Hidalgo23, Hsin23, Thomay23, Botha24, Biwas24, Wimmer24} & Hardware & Medium & High & High \\[3pt]

FOV Restriction & \citep{Alzayer19, Groth21, Shi21, Oh22, Won22} & Software & High & Low & Low \\
Dynamic Speed Control & \citep{Oh22, Souchet23b, Botha24} & Software & High & Low & Low \\
Fixed Elements & \citep{Oh22, Souchet23b, Biwas24, Botha24} & Software & High & High & High \\
Locomotion Type & \citep{Rahimi18} & Software & High & Low & Low \\
Peripheral Blurring & \citep{Nie19, Hussain21} & Software & Medium & Low & Low \\
Reverse Optical Flow & \citep{Kim23} & Software & Medium & Low & Low \\
Dynamic Depth-of-Field & \citep{Souchet23b, Botha24} & Software & Medium & Medium & Medium \\
Simulation Intensity Modification & \citep{Oh22, Biwas24, Botha24} & Software & High & Low & Low \\
Interpupillary Distance Calibration & \citep{Stanney20, Cross22} & Software & High & High & High \\
Limiting Exposure Duration & \citep{Cross22, Souchet23b, Biwas24} & Software & High & Low & Medium \\
Simulation Evaluation Framework & \citep{Lee24, Oberhauser18} & Software & Medium & High & Medium \\[3pt]

Postural Instructions \& Monitoring & \citep{Litleskare21, Tian22, Souchet23a, Biwas24, Botha24} & Ergonomic & High & High & High \\
HMD Weight Management & \citep{Chen21, Souchet23b} & Ergonomic & Medium & High & High \\
Adjusting Required Movements & \citep{Kim20b, Souchet23a, Botha24} & Ergonomic & High & Low & Low \\[3pt]

Pleasant Music & \citep{Keshavarz14, Kourtesis23, Rahimzadeh23} & Psych & High & Low & Low \\
Onboarding with Positive Framing & \citep{Mao21, Chauvergne23, Cross22, Biwas24} & Psych & High & High & High \\
Screening (Questionnaires) & \citep{Oberhauser18, Kim20a, Alashwal21, Laudien22, Tian22, Moesl23, Biwas24} & Psych & High & High & High \\[3pt]

Diaphragmatic Breathing & \citep{Russell14, Stromberg15} & Physio & High & Medium & Medium \\
Airflow & \citep{Damour17, Paroz21, Matviienko22} & Physio & High & High & High \\
Mastication Effects & \citep{Kaufeld22} & Physio & High & Medium & Medium \\
Galvanic Vestibular Stimulation & \citep{Sra19, Weech20, Groth22} & Physio & Low & Medium & Low \\
Habituation & \citep{Gavgani17, Kroma21, Wienrich22, Rahimzadeh23} & Physio & Medium & High & High \\
Oculomotor Training & \citep{Kim18, Park17} & Physio & Medium & High & High \\
Pharmacological Intervention & \citep{Jarisch14, Zaghlool15, Rahimzadeh23} & Physio & Medium & High & Medium \\

\end{longtable}
\normalsize
} 

\subsubsection{Individual factors}
Individual factors influence how pilots physiologically and psychologically respond to MR HMD-based training environments, affecting susceptibility to cybersickness, fatigue, discomfort, and suboptimal cognitive workload. Evidence in this area is derived predominantly from VR-based studies, although many findings are likely transferable to MR contexts due to shared HMD-related perceptual and sensorimotor mechanisms. Within helicopter pilot training, these factors are best understood as moderating susceptibility to adverse effects rather than independently determining training effectiveness, with their operational significance often shaped by broader simulation, hardware, and fidelity-related constraints.

\paragraph*{Age}  
Age-related declines in visual accommodation and vestibular sensitivity are associated with increased vulnerability to VIMS and cybersickness \citep{Shafer17,Tian22}. Across immersive VR and MR studies, older participants generally report higher Simulator Sickness Questionnaire (SSQ) and Virtual Reality Symptom Questionnaire (VRSQ) scores than younger users \citep{Rebenitsch14,Hildebrandt18,Oh22}. However, the practical relevance of age in MR helicopter pilot training may be comparatively limited, as aviation populations typically exclude pediatric and geriatric cohorts through medical certification requirements. Most trainees are young to middle-aged adults with relatively preserved sensorimotor function, although mid-career pilots may still experience age-related changes such as presbyopia or slower vestibular adaptation. Consequently, while age appears to be a consistent individual risk factor across MR/VR literature, current evidence suggests it is likely to play a secondary role in operational MR helicopter training contexts relative to factors such as latency, visual motion, and simulator fidelity.

\paragraph*{Gender}  
Evidence regarding gender differences in human-factors and MR-HMD usability remains mixed and is likely multifactorial rather than attributable to a single cause. Some studies report greater nausea or discomfort among female participants during fast-moving VR and MR tasks \citep{Tian22,Jasper23,Kourtesis23,Biwas24}, whereas others observe no significant gender effects once headset calibration factors such as interpupillary distance (IPD) are appropriately adjusted \citep{Curry20}. Although anatomical differences in IPD are frequently discussed in the literature, with women on average exhibiting smaller IPDs than men \citep{Stanney20}, IPD mismatch alone is unlikely to fully explain observed gender-related differences, particularly given that motion-sickness susceptibility has also been reported in non-HMD environments. Instead, current evidence suggests that visual, vestibular, ergonomic, and physiological factors may interact to influence individual susceptibility. Beyond optical alignment, ergonomic considerations may also contribute to differential discomfort. HMD weight and load distribution can place strain on the cervical spine and neck musculature during prolonged use, with ergonomic effects potentially varying across pilots due to anthropometric differences in body size, neck strength, and cockpit fit requirements. This is particularly relevant in MR helicopter pilot training, where sustained head movements, visual scanning, and repeated cross-cockpit instrument referencing are operationally necessary. EASA FSTD Special Conditions for HMD-based helicopter simulation further reinforce the importance of maintaining accurate eyepoint alignment, visual clarity, and stable visual presentation during cockpit operations \citep{EASA23}.  

Accordingly, mitigation strategies related to IPD calibration and HMD weight management appear particularly important for maintaining comfort and usability across users. Proper IPD calibration is similarly critical for preserving visual alignment and minimizing eye strain during long-duration training sessions, with modern HMDs typically supporting built-in calibration procedures and EASA FSTD conditions explicitly requiring accurate eyepoint and projection geometry calibration \citep{Stanney20,Cross22,EASA23}. In parallel, ergonomic strategies such as improved head-strap support and balanced weight distribution may help reduce discomfort without compromising simulator fidelity \citep{Chen21,Souchet23b,Plumer23}. Among these approaches, IPD calibration demonstrates the strongest feasibility and contextual suitability for MR helicopter pilot training because it can generally be implemented through existing hardware-selection and configuration procedures while remaining compatible with EASA-oriented fidelity requirements, resulting in a high overall priority rating. Although HMD weight-management strategies appear operationally valuable, supporting evidence remains derived predominantly from VR-based rather than MR-specific aviation studies. More broadly, current evidence regarding gender-specific susceptibility remains constrained by heterogeneous findings, convenience sampling, and limited MR aviation validation, suggesting that these mitigation strategies should presently be interpreted as broadly improving ergonomic inclusivity and user comfort rather than addressing definitively established gender-specific limitations.

\paragraph*{Motion-sickness susceptibility}  
Motion-sickness susceptibility refers to an individual’s predisposition to experience motion-related symptoms such as nausea, dizziness, disorientation, and discomfort when exposed to sensory-motion stimuli. It is commonly assessed using instruments such as the Motion Sickness Susceptibility Questionnaire (MSSQ) or through documented histories of sea-, air-, or simulator-sickness experiences. Individuals with high baseline susceptibility are more likely to experience discomfort, nausea, and disorientation during immersive HMD exposure \citep{Oh22,Jasper23}. In MR helicopter pilot training, this susceptibility is operationally important because simulated aircraft motion, head movement, and cockpit visual flow may conflict with vestibular and proprioceptive cues, potentially disrupting concentration during high-precision flight tasks. Although most evidence is derived from VR rather than MR-specific aviation studies, the underlying sensory-integration mechanisms are transferable to MR HMD training contexts because both modalities rely on visually mediated motion cues.

The most practical first-line strategies are screening, limiting exposure duration, and habituation. Standardized questionnaires are highly feasible, contextually suitable, and high priority because they can be used before or after training to identify susceptibility, workload, and cybersickness risk without altering simulator fidelity \citep{Oberhauser18,Kim20a,Alashwal21,Laudien22,Tian22,Biwas24}. However, questionnaires are limited by their retrospective and subjective nature, meaning they are better suited to screening and evaluation than real-time intervention. Habituation is also contextually suitable and high priority, as gradual exposure can improve tolerance to visually induced motion over repeated sessions \citep{Gavgani17,Kroma21,Wienrich22,Rahimzadeh23}. Its feasibility is rated as medium because it requires integration into existing lesson structures and may extend training schedules. Similarly, limiting exposure duration is highly feasible and may benefit motion-sensitive trainees \citep{Cross22,Biwas24,Souchet23b}, but its context suitability is low because strict time limits may conflict with the duration and realism expected in helicopter simulator training. Simulation intensity modification may be useful during onboarding\citep{Oh22,Biwas24,Botha24}, but because reducing visual motion or complexity can undermine fidelity, it is best treated as a low-priority strategy for core MR helicopter training.

Biometric monitoring offers a promising but more complex supplementary strategy. Measures such as electroencephalography (EEG), eye tracking, pupillometry, heart rate variability (HRV), and galvanic skin response (GSR) can provide insight into workload, discomfort, affective state, and emerging cybersickness responses \citep{Luong20,Krokos22,Hidalgo23,Hsin23,Thomay23,Wimmer24}. The strength of evidence varies by modality: EEG, pupillometry, and eye-tracking measures are comparatively stronger for workload, attention, and error-related monitoring, whereas GSR is more vulnerable to confounding because sympathetic arousal can reflect stress, workload, cognitive demand, or emotional activation rather than cybersickness specifically. This is especially important in helicopter training, where elevated arousal may arise from the task itself rather than from HMD-induced discomfort. Accordingly, biometric monitoring is rated as high in context suitability and priority because pilot-state monitoring is highly relevant to safety-critical training, but only medium in feasibility because attaching complex sensors to each trainee can increase setup burden, reduce comfort, and interfere with training flow. Based on the RoB profile, support for biometric approaches is mixed: some studies show stronger methodological quality, such as workload classification in VR piloting \citep{Luong20}, while other relevant MR or flight-related applications remain limited by higher RoB, smaller validation bases, or non-operational contexts \citep{Hidalgo23,Tichon14}. Therefore, biometric monitoring should currently be interpreted as an adjunct to questionnaires and instructor observation rather than a standalone trigger for adaptive cybersickness mitigation.

Pharmacological and supplement-based interventions may provide targeted support for individuals with high motion-sickness susceptibility, as cybersickness and motion sickness share similar physiological pathways \citep{Mazloumi18}. Interventions including antihistamines, antimuscarinics, ginger, and vitamin C have been explored in relation to motion sickness and VIMS \citep{Jarisch14,Zaghlool15,Rahimzadeh23}. However, these approaches remain supplementary rather than primary strategies in MR helicopter training due to potential side effects, prescribing considerations, and operational constraints, resulting in medium feasibility and priority despite relatively high contextual suitability. Mastication effects may represent a simpler and more practical physiological intervention. \citet{Kaufeld22} reported that chewing gum during a VR helicopter simulation reduced VIMS, with findings suggesting that the mechanical act of chewing itself was the primary contributor. This strategy is rated high in feasibility and medium in context suitability and priority because it is inexpensive, non-invasive, and operationally simple to implement \citep{Kaufeld22,Biwas24,Botha24}. Nevertheless, individual user preference and the practicality of chewing gum during simulator sessions may limit widespread adoption. The evidence supporting mastication effects is promising but currently limited to a small number of controlled VR studies with moderate RoB and limited MR-specific validation. Although such interventions may be contextually suitable for carefully screened and consenting trainees, they are not a first-line strategy for MR helicopter training. Overall, susceptibility-related mitigation should prioritize screening and habituation, with biometric monitoring and pharmacological/mastication approaches used only as supplementary or targeted supports where operationally justified.

\paragraph*{Physical condition}  
A pilot's physical and neurological condition influences how effectively sensory cues from an MR HMD are tolerated and integrated during training. Factors such as heightened body awareness, poor sleep, acute fatigue, minor illness, migraine, uncorrected vision problems, or vestibular/neurological conditions may increase vulnerability to discomfort, nausea, disorientation, and reduced task focus \citep{Tian22,Biwas24}. In MR helicopter pilot training, this is particularly important because trainees must sustain visual scanning, head movement, cockpit interaction, and high attentional control while exposed to simulated aircraft motion. Although severe medical or neurological conditions are typically screened through aviation medical certification, day-to-day fluctuations such as sleep debt, stress, or minor illness remain operationally relevant and should be captured through pre-simulation checks. Accordingly, previously discussed strategies such as pre-simulation screening, selective biometric monitoring, and habituation may provide additional support for identifying and managing transient physical-state risks before symptoms substantially affect training performance.

Beyond these broader screening and monitoring approaches, three additional mitigation strategies are particularly relevant to supporting pilots’ immediate physiological stability during MR training: postural instructions and monitoring, airflow, and diaphragmatic breathing. Postural instructions and monitoring are rated as high feasibility, high context suitability, and high priority because they can be integrated into standard simulator setup without compromising fidelity. Guidance on seated posture, HMD fit, seat position, and control alignment can reduce neck and spinal strain, while instructor feedback or HMD-based spatial tracking can support correction during training \citep{Litleskare21,Tian22,Souchet23a,Botha24,Biwas24}. This is especially compatible with helicopter simulation because EASA-oriented training already depends on maintaining accurate pilot eyepoint, cockpit alignment, and representative control positioning \citep{EASA12,EASA23}. The supporting evidence is reasonably consistent but mixed in quality: empirical work on postural stability is moderate in RoB \citep{Litleskare21}, whereas broader ergonomic recommendations often come from review or framework papers with lower review-level confidence, meaning the strategy is practically strong but still requires more MR aviation-specific validation \citep{Souchet23a,Botha24,Biwas24}.

Airflow is also rated as high feasibility, high context suitability, and high priority because it is simple to implement, non-invasive, and does not conflict with simulator realism. Directed airflow may reduce disorientation and nausea by providing cooling and sensory grounding, while also resembling cockpit ventilation in aviation contexts \citep{Damour17,Paroz21,Matviienko22}. Its evidence base is comparatively stronger than many physiological strategies: \citet{Damour17} and \citet{Matviienko22} are empirical studies with low-to-moderate RoB, while \citet{Paroz21} provides aviation-relevant VR pilot-navigation evidence with moderate RoB. As a result, airflow appears to be one of the more operationally credible comfort-support strategies for MR helicopter training, particularly for trainees experiencing transient discomfort related to physical state. Furthermore, diaphragmatic breathing is rated as high feasibility, medium context suitability, and medium priority. It is low-cost and can regulate autonomic arousal through slow, controlled breathing, with evidence from controlled studies suggesting benefits for motion-sickness reduction \citep{Russell14,Stromberg15}. However, its suitability during active helicopter training is more limited because maintaining breathing technique may impose additional cognitive demand and distract from flight tasks. Therefore, diaphragmatic breathing is best positioned as a pre-training regulation strategy or recovery technique during breaks rather than as an active in-task intervention. Both supporting empirical studies show relatively low RoB, but they are not MR aviation studies, so transfer to MR helicopter training should be interpreted cautiously. Overall, physical-condition-related mitigation should prioritize simulator setup, postural instructions/monitoring, and airflow as practical first-line strategies, with breathing-based regulation used selectively when it does not interfere with task performance.

\paragraph*{Emotional condition}  
Emotional condition substantially influences how pilots respond to immersive MR training environments, particularly through its interaction with stress, workload, attentional focus, and cybersickness susceptibility. Negative emotional states such as anxiety, frustration, stress, or techno-stress can elevate sympathetic arousal and increase attention toward bodily discomfort, thereby amplifying VIMS and cybersickness symptoms \citep{Tian22,Biwas24,Souchet23a}. Higher perceived workload has also been associated with elevated SSQ scores in immersive environments \citep{Jasper23}. Within MR helicopter pilot training, these effects are operationally significant because pilots must maintain situational awareness, decision-making, and communication performance under cognitively demanding conditions. Emotional dysregulation may therefore degrade non-technical skills essential to aviation safety, including problem-solving, attention allocation, and situational awareness \citep{Tichon14}. 

The most operationally suitable mitigation strategy identified in the literature is onboarding with positive framing, which is rated as high feasibility, high context suitability, and high priority. Effective onboarding includes clear headset-fitting guidance, posture correction, calibration support, and familiarization with the MR environment \citep{Chauvergne23,Cross22}. Importantly, positively framed onboarding may also reduce nocebo-related symptom amplification by avoiding excessive emphasis on cybersickness risk. \citet{Mao21} demonstrated that negative cybersickness framing can substantially increase symptom reporting, highlighting how expectation effects can shape immersive experiences independently of hardware or simulation fidelity. In aviation contexts, this is particularly relevant because anxiety or anticipatory discomfort may interfere with concentration during high-workload flight tasks. Accordingly, onboarding procedures that promote confidence, comfort, and task familiarity appear well aligned with MR helicopter pilot training requirements while preserving immersion and simulator realism. The supporting evidence base is moderately strong overall: \citet{Mao21} and \citet{Jasper23} are empirical studies with moderate-to-low RoB, while \citet{Chauvergne23} contributes mixed empirical and interview-based evidence regarding onboarding practices in immersive systems. However, much of this literature remains VR-based and not specific to operational aviation contexts. 

Pleasant music has also been explored as a potential psychological mitigation strategy because calming auditory stimuli may reduce stress and discomfort responses \citep{Keshavarz14,Kourtesis23,Rahimzadeh23}. However, despite high feasibility, its context suitability and priority are low for MR helicopter pilot training. Music may interfere with cockpit communications, auditory cue processing, and training realism, all of which are critical in aviation environments. Consequently, while pleasant music may have value in general VR comfort studies, it appears operationally unsuitable for most helicopter simulation scenarios. Furthermore, the supporting evidence is comparatively heterogeneous, with studies differing substantially in task context, immersion type, and outcome measures. Overall, emotional-condition-related mitigation in MR helicopter pilot training should primarily prioritize structured onboarding with positive framing rather than relaxation-oriented interventions that may conflict with operational fidelity.

\subsubsection{Hardware factors}
Hardware factors refer to the technical capabilities of the MR HMD and its supporting visual system. In MR helicopter pilot training, these factors are especially important because the HMD must support both comfort and fidelity: pilots need to read instruments, perceive external visual cues, and maintain stable visual alignment during dynamic cockpit tasks. Many hardware-related findings are derived from VR studies, but their relevance transfers strongly to MR because display resolution, latency, tracking stability, optical geometry, and rendering quality are shared constraints across HMD-based systems.

\paragraph*{Display resolution}  
Display resolution affects both visual comfort and the fidelity of simulated cockpit information. Higher-resolution displays improve visual clarity, reduce pixelation, and support more accurate perception \citep{Wang22} of cockpit instruments and external visual cues. This is particularly important in MR helicopter pilot training because trainees must read fine-grained instrument symbology while maintaining awareness of external scene features and cross-cockpit cues. Evidence from \citet{Wang22} suggests that increasing resolution from low-resolution displays to higher-resolution displays can reduce cybersickness and visual discomfort, although benefits may plateau once sufficient clarity is reached. This finding comes from a controlled VR study with moderate RoB, meaning the measurement basis is useful but still requires cautious transfer to MR aviation contexts. The most relevant mitigation strategy is HMD choice, which is rated as high feasibility, high context suitability, and high priority. From a feasibility perspective, selecting an HMD with sufficient resolution is primarily a procurement and configuration decision rather than a major change to training procedure. From a context-suitability perspective, high-resolution displays preserve rather than compromise simulator fidelity, making them well aligned with helicopter training needs. EASA FSTD Special Conditions reinforce this priority by requiring objective testing of visual-system quality, including vernier resolution, grating resolution, and system geometry from the pilot eye-point \citep{EASA23}. These requirements indicate that display clarity is not merely a comfort issue, but part of the qualification logic for HMD-based helicopter simulation. 

\paragraph*{Latency and latency jitter}  
Latency, defined as the delay between pilot head/control movement and the corresponding visual update, is a primary hardware driver of VIMS and cybersickness in HMD-based simulation. Increased display lag has been shown to increase scene-instability ratings, reduce spatial presence, and heighten cybersickness symptoms \citep{Kim20b}, with related concerns also identified across VR training and review literature \citep{Tian22,Souchet23b,Botha24}. Closely related is latency jitter, which refers to frame-to-frame variation around the mean delay. Even when average latency remains within acceptable thresholds, irregular timing can degrade motion predictability and introduce micro-stutters during rapid head movements, intensifying visual–vestibular mismatch and discomfort \citep{Stauffert18}. In MR helicopter pilot training, these issues are particularly critical because precise sensorimotor coupling supports instrument scanning, external visual referencing, and control timing. EASA FSTD Special Conditions therefore require combined sensor-plus-render latency to remain $\leq$20 ms for Flight Training Device (FTD) Level 3 qualification \citep{EASA23}. The main mitigation strategy is HMD choice again. Selecting an HMD with low motion-to-photon latency, minimal latency jitter, stable tracking, and sufficient rendering performance is feasible and is highly suitable because it improves comfort while preserving simulator fidelity. The empirical evidence for latency effects is reasonably strong but remains predominantly VR-based. \citet{Kim20b} and \citet{Stauffert18} provide controlled experimental evidence with moderate-to-high RoB, while broader support from reviews and technical papers contributes contextual support but lower direct evidential strength. However, due to strict regulatory requirements, both mean latency and latency stability should be treated as high-priority hardware requirements for MR helicopter training, with further MR-specific aviation validation still required.

\paragraph*{Type of display}  
The type and quality of the HMD display subsystem strongly influence visual comfort, immersion, and cybersickness in MR environments. High-end displays incorporating advanced LED panels or high-grade LCD optics provide improved color accuracy, contrast, clarity, and refresh stability, contributing to a more visually stable cockpit scene and reduced visual fatigue \citep{Yadin18,Wang21,Biwas24}. In contrast, lower-quality displays are more prone to visual artifacts, reduced clarity, and inconsistent refresh behavior, which may increase discomfort and cybersickness \citep{Tian22}. In MR helicopter pilot training, display quality is especially important because pilots must continuously interpret cockpit instruments, terrain, and external flight cues under high attentional demand. EASA FSTD Special Conditions further reinforce this requirement through objective standards for frame rate, system geometry, color representation, vernier and grating resolution, chromatic aberration, IPD calibration, and 3D projection accuracy \citep{EASA23}. The primary mitigation strategy is HMD choice, as selecting a high-end HMD with stable high-refresh displays, accurate optics, and integrated IPD calibration is operationally feasible and directly supports both comfort and simulator fidelity requirements. 

Some advanced systems additionally incorporate dynamic depth-of-field mechanisms, such as gaze-driven autofocus, to reduce VAC and visual strain \citep{Varjo}. Dynamic depth-of-field is rated medium in feasibility, context suitability, and priority because software-based blur manipulation may interfere with cockpit scanning and environmental awareness if poorly implemented. However, hardware-integrated focal-adjustment systems appear more promising because they are specifically optimized for naturalistic depth rendering in immersive tasks \citep{Souchet23b,Botha24}. The evidence supporting display-related human factors is derived primarily from VR studies and review literature, with comparatively limited MR-specific aviation validation. Empirical evidence regarding refresh stability and visual discomfort is generally moderate in strength, while many broader recommendations originate from technical and review papers with lower direct evidential rigor. Nevertheless, the consistency between empirical findings, review synthesis, and EASA visual-system requirements suggests that high-quality display systems are a critical hardware consideration for MR helicopter pilot training.

\paragraph*{Type of input}  
The type of input used to interact with the MR environment influences both usability and cybersickness risk. Input systems requiring exaggerated or unnatural body movements, such as gesture-based controls, can increase visual–vestibular conflict and physical strain, thereby elevating cybersickness likelihood \citep{Mittelstaedt18}. By contrast, stable and familiar input methods, such as joysticks and physical cockpit controls, generally produce lower cybersickness and improve interaction precision \citep{Monteiro20}. In MR helicopter pilot training, this distinction is particularly important because pilots must perform accurate control inputs under high workload while maintaining situational awareness. EASA FTD Level 3 standards further require a full-scale cockpit replica with representative helicopter controls, making physical aviation controls a practical necessity rather than an optional design feature \citep{EASA12}. 

The primary mitigation strategy is input type, rated high in feasibility, context suitability, and priority. Utilizing real or replica helicopter controls aligns closely with operational cockpit workflows, reduces unnecessary body movement, and lowers cognitive overhead associated with unfamiliar interaction methods \citep{Cross22,Kimura24}. Physical input systems additionally improve realism and tactile consistency compared with purely virtual controls \citep{Laudien22}. Haptic feedback may further support tactile realism \citep{Cross22,Kimura24}, but in EASA-oriented helicopter simulators, full-scale physical cyclic and collective controls already provide the most contextually appropriate tactile cues, making separate haptic devices a low priority. Consequently, haptic feedback is rated medium in feasibility due to additional hardware requirements and low in contextual suitability for helicopter simulation. The supporting evidence is moderately strong but largely VR-derived: empirical studies examining controller type and motion-related discomfort generally report moderate RoB, while aviation-focused MR evidence remains comparatively limited. Nevertheless, the convergence between empirical findings and EASA cockpit-replication requirements strongly supports the use of physical aviation controls as a key mitigation strategy for human-factor challenges in MR helicopter pilot training.

\subsubsection{Simulation content factors}
Simulation content factors refer to the design and functionality of the MR environment itself, including how the virtual world is structured and presented. These factors influence the realism and immersion experienced by the pilots, affecting how well they can engage with the training content. Although much of the evidence is derived from VR studies, many findings are likely transferable to MR due to shared HMD-mediated visual and sensorimotor mechanisms.

\paragraph*{Visual motion}  
Visual motion is a major contributor to cybersickness in immersive MR environments. High translational speeds, rotational movements, rapid scene transitions, and strong peripheral motion cues can intensify vection and increase visual--vestibular conflict \citep{Oh22,Botha24,Biwas24}. This is especially relevant in MR helicopter pilot training, where flight maneuvers inherently involve continuous motion, banking, altitude changes, and external scene movement. Several mitigation strategies have been proposed, but their suitability differs sharply in aviation contexts. Dynamic speed control, peripheral blurring, and reverse optical flow may reduce cybersickness by suppressing or altering perceived motion \citep{Nie19,Hussain21,Kim23}, but are rated low in context suitability and priority because they may compromise peripheral awareness, motion realism, and training transferability. Dynamic speed control is rated high in feasibility, while peripheral blurring and reverse optical flow are rated medium in feasibility, because these approaches can generally be implemented through software-level modifications without substantial hardware changes. 

Galvanic vestibular stimulation (GVS) may also reduce sensory conflict by modulating vestibular input \citep{Groth22,Sra19,Weech20}, but is rated low in feasibility and priority due to hardware complexity and limited operational readiness, while only demonstrating medium contextual suitability because of its limited compatibility with conventional simulator workflows. Motion platforms may reduce sensory conflict by aligning visual motion with vestibular cues, but their low feasibility due to cost, space, and maintenance constraints limits priority despite high contextual suitability for helicopter simulation \citep{Kim20a,EASA12}. Overall, the evidence for visual-motion mitigation is largely derived from controlled VR studies rather than MR aviation environments, with most empirical studies showing moderate RoB. The relevant motion-platform evidence from \citet{Kim20a} was rated as high RoB, mainly limiting confidence in the strength of its empirical claims. Therefore, although these strategies can reduce symptoms experimentally, preserving realistic motion representation while improving system stability appears more suitable for MR helicopter training than aggressively suppressing motion cues.

\paragraph*{Means of locomotion}  
The way users move through an immersive environment influences both cybersickness susceptibility and perceived realism. Continuous locomotion methods involving body-based or hand-controlled movement can increase sensory conflict and cybersickness because users experience visual motion without corresponding physical movement \citep{Rahimi18}. Teleportation-based locomotion generally reduces discomfort by removing continuous visual motion, but may also reduce presence and embodiment within the environment \citep{Rahimi18,Botha24}. In MR helicopter pilot training, this trade-off is particularly important because realistic continuous motion perception is fundamental to flight maneuvering, spatial orientation, and training transferability \citep{Southgate20}. The primary mitigation strategy is altering locomotion type, specifically to teleportation-based movement, which is rated high in feasibility but low in context suitability and priority. While teleportation effectively reduces cybersickness in immersive systems \citep{Rahimi18}, it disrupts the continuity of motion and spatial navigation required for realistic helicopter simulation. Consequently, although teleportation may be suitable in general VR navigation tasks, it is poorly aligned with aviation training objectives where sustained motion cues and environmental continuity are operationally essential. The supporting evidence is derived primarily from controlled VR studies with moderate RoB, while MR-specific and aviation-specific evidence remains limited. Overall, preserving realistic continuous locomotion appears more appropriate for MR helicopter pilot training despite the associated cybersickness challenges.

\paragraph*{Field of view}  
FOV strongly influences immersion, spatial awareness, and cybersickness within MR environments. Wider FOVs improve embodiment, depth perception, and environmental awareness by more closely approximating natural human vision \citep{Cross22,Southgate20}. In aviation contexts, these properties are particularly important because pilots rely heavily on peripheral vision and cross-cockpit awareness during flight tasks. Some studies further suggest that sufficiently wide FOVs may reduce nausea by enhancing immersion and visual consistency \citep{Fussell24}. However, excessively wide FOVs can also increase visual-processing demands and intensify cybersickness through stronger vection effects \citep{Oh22,Tian22,Souchet23b}. Conversely, overly restricted FOVs may reduce discomfort but can impair realism, presence, and situational awareness, while also increasing compensatory head movements. EASA FSTD Special Conditions therefore specify minimum cross-cockpit FOV requirements of $\pm40^\circ$ horizontally and $30^\circ$ up and $35^\circ$ down vertically to preserve realistic training conditions \citep{EASA23}. 

Two primary mitigation strategies emerge: FOV restriction and HMD choice. FOV restriction is rated high in feasibility but low in context suitability and priority. Although narrowing the visible scene can reduce cybersickness \citep{Alzayer19,Groth21,Shi21,Oh22,Won22}, doing so conflicts with the fidelity and peripheral-awareness requirements of helicopter pilot training. In contrast, HMD choice is rated high in priority because selecting an HMD with a sufficiently wide, stable, and optically consistent FOV supports both immersion and regulatory-aligned realism. This is especially important for maintaining cockpit awareness and minimizing compensatory head movement during flight tasks. The supporting evidence is primarily VR-based, with empirical studies on FOV restriction generally demonstrating moderate RoB, while broader recommendations regarding immersion and simulator realism are supported by reviews and technical papers with lower direct evidential strength. Overall, preserving an appropriately wide and stable FOV appears substantially more suitable for MR helicopter pilot training than artificially restricting visual input to reduce cybersickness.

\paragraph*{Type of content}  
The type and intensity of simulated content strongly influence cybersickness, workload, and immersion within MR environments. Fast-moving scenes, strong peripheral motion, rapid environmental transitions, uneven terrain, and adverse weather conditions have all been associated with increased discomfort and sensory conflict \citep{Oh22,Keshavarz19,Ang22,Alashwal21}. In helicopter pilot training, these effects are particularly relevant because realistic flight simulation inherently involves dynamic motion, environmental variability, and complex visual processing demands. While reducing simulation intensity may alleviate discomfort, doing so risks compromising fidelity and training transferability. Accordingly, simulation intensity modification is rated high in feasibility but low in context suitability and priority, as simplifying flight dynamics or environmental complexity conflicts with the objective of reproducing realistic helicopter operations \citep{Oh22,Biwas24,Botha24}. 

In contrast, fixed visual elements are rated high in feasibility, context suitability, and priority. Stable cockpit structures and instrumentation may naturally reduce sensory conflict without substantially disrupting immersion, unlike artificial overlays such as crosshairs or virtual noses \citep{Oh22,Souchet23b,Biwas24,Botha24}. Simulation evaluation frameworks have also emerged as a promising strategy for balancing realism and comfort. \citet{Lee24} proposed a structured assessment framework integrating hardware, content, and user-related factors to evaluate cybersickness, comfort, and excitement across immersive environments. As modern helicopter simulators already capture telemetry and performance data, integrating such frameworks into MR helicopter training appears operationally feasible without major system redesign \citep{Oberhauser18}. Accordingly, these frameworks demonstrate high contextual suitability but only medium priority because they primarily support evaluation and optimization rather than directly mitigating symptoms. Across these strategies, the evidence base remains predominantly VR-derived and of moderate methodological quality, with empirical findings generally showing moderate RoB and many implementation recommendations originating from review or framework papers rather than operational MR aviation studies.

\paragraph*{Exposure duration}  
Prolonged exposure to MR environments increases the likelihood of fatigue, visual discomfort, and cybersickness due to sustained sensory and cognitive demands \citep{Tian22,Souchet23a}. In helicopter pilot training, this is particularly important because training sessions are often extended and cognitively demanding, requiring sustained attention and visuospatial processing. Although shorter sessions and structured breaks can reduce discomfort and preserve cognitive performance, strict exposure limits may conflict with operational realism and the extended durations characteristic of conventional flight training. Accordingly, limiting exposure duration is rated high in feasibility but low in context suitability and medium in overall priority. While short sessions reduce symptoms, strict duration caps may conflict with extended helicopter training requirements. Habituation-based approaches therefore appear more operationally suitable. Gradual exposure protocols aim to improve tolerance to immersive motion stimuli over time, reducing discomfort during longer sessions \citep{Rahimzadeh23,Gavgani17,Kroma21,Wienrich22}. While implementation requires adjustments to training schedules and onboarding procedures, habituation is rated high in context suitability because it preserves simulator fidelity rather than reducing it. Across these strategies, the supporting evidence is primarily derived from controlled VR studies with generally moderate methodological quality and limited MR-specific aviation validation. Nevertheless, the consistency of findings across multiple experimental studies and review papers suggests that gradual adaptation strategies are likely transferable to MR helicopter training contexts.

\paragraph*{Visual complexity}  
Dense and highly detailed virtual scenes increase cognitive load and sensory conflict, elevating cybersickness and fatigue whilst reducing comfort during demanding MR tasks \citep{Souchet23a,Biwas24,Botha24}. In helicopter pilot training, however, high visual fidelity is also necessary to support situational awareness, instrument interpretation, and realistic environmental perception. Consequently, MR simulation design requires a balance between visual realism and usability. Evidence from \citet{Deluca23} suggests that usability and interface clarity contribute more strongly to perceived realism and immersion than visual learnability alone, indicating that effective interface design may improve comfort without substantially reducing fidelity. Lowering visual complexity through simulation intensity modification is therefore rated high in feasibility but low in context suitability and priority, reflecting the risk of reducing flight realism and training transferability. Peripheral blurring has similarly been explored as a method for reducing visual overload and motion sensitivity \citep{Nie19,Hussain21}, but is of low context suitability for aviation because pilots rely heavily on clear peripheral awareness for monitoring cockpit instruments and spatial orientation. Across these strategies, the supporting evidence is largely derived from VR studies of moderate methodological quality, with limited MR-specific aviation validation. Nevertheless, findings suggest that optimizing interface usability and reducing unnecessary visual clutter may be more operationally appropriate than reducing simulation fidelity in MR helicopter training.

\subsubsection{Ergonomic factors}
Ergonomic factors relate to the physical comfort and biomechanical demands associated with MR HMD use during training. Because VR and MR headsets share highly similar form factors, weight distribution systems, and head-mounted interaction requirements, ergonomic findings from VR studies are considered broadly transferable to MR aviation contexts. These factors directly influence fatigue, usability, and sustained training performance during prolonged simulation exposure.

\paragraph*{HMD weight}  
HMD weight is a major ergonomic contributor to physical fatigue and discomfort, with heavier devices increasing musculoskeletal strain on the neck, shoulders, and upper back during extended use \citep{Souchet23a}. This is particularly relevant in helicopter pilot training, where frequent head movements and prolonged visual scanning are operationally necessary. While lighter HMDs generally improve comfort, reducing device weight may require trade-offs in display quality, tracking capability, or other high-fidelity features important for aviation simulation. Accordingly, HMD choice and HMD weight management are both rated high in context suitability and priority. Selecting high-end HMDs that balance ergonomic comfort with technical capability is critical for maintaining training realism while minimizing physical strain \citep{Cross22,Souchet23b,Biwas24,Botha24}. However, because advanced systems often increase overall headset mass, ergonomic mitigation strategies such as improved strap systems or hanging support mounts may be required to redistribute load during longer sessions \citep{Chen21,Plumer23}. Consequently, HMD weight management strategies demonstrate medium feasibility due to additional equipment and setup requirements, despite high contextual suitability and priority. These approaches are operationally compatible with MR helicopter simulators and align with EASA requirements for high-performance HMD-based training devices \citep{EASA23}. The supporting evidence is primarily derived from VR ergonomic and usability studies with generally moderate methodological quality, but findings are consistent across both empirical and review-level literature, supporting their likely transferability to MR aviation training environments.

\paragraph*{Required movements}  
Required physical movements within MR environments contribute to both ergonomic strain and cybersickness, particularly during prolonged or repetitive interaction sequences. Frequent reaching actions, repeated cockpit interactions, and rapid head rotations can increase muscle fatigue and discomfort over time \citep{Souchet23a}. Additionally, \citet{Kim20b} reported greater scene instability and cybersickness during faster head oscillation rates, suggesting that rapid or repeated head movements intensify sensory conflict and physical strain. In helicopter pilot training these movements are often unavoidable because pilots must continuously scan instruments, monitor the external environment, and interact with cockpit controls to replicate realistic flight procedures. Accordingly, adjusting required movements is rated high in feasibility but low in context suitability and priority. Although reducing repetitive motions or minimizing head rotations may improve ergonomic comfort in general VR/MR applications \citep{Kim20b,Souchet23a,Botha24}, doing so would compromise procedural realism and negatively affect training transferability in aviation contexts. The supporting evidence is primarily derived from controlled VR studies and review literature with generally moderate RoB, but findings remain relevant to MR helicopter training due to the similar head-mounted interaction demands shared across VR and MR systems.

\paragraph*{Posture}  
Posture plays an important role in both ergonomic comfort and cybersickness during MR exposure. Seated postures generally produce lower cybersickness and greater physical stability than standing configurations \citep{Biwas24}, which is particularly relevant for helicopter pilot training where users remain seated within a cockpit environment for extended periods. Poor neck, head, or upper-body posture can increase musculoskeletal strain and accelerate fatigue, reducing comfort and attentional performance during training \citep{Tian22,Souchet23a,Botha24}. Studies examining postural stability further suggest that cybersickness is associated with greater deterioration in balance and body control during immersive exposure \citep{Litleskare21}. Accordingly, postural instructions and monitoring are rated high in feasibility, context suitability, and priority. Ergonomic guidance relating to seated posture, HMD fit, and cockpit alignment may reduce strain and improve stability during prolonged MR sessions \citep{Litleskare21,Tian22,Souchet23a,Botha24,Biwas24}. These strategies align well with EASA-oriented cockpit interaction requirements and may be further supported through instructor feedback or headset-tracking data \citep{EASA23}. Although the evidence base is derived primarily from VR ergonomic studies with generally moderate RoB, findings are relatively consistent across empirical and review literature, supporting likely transferability to MR helicopter training.

\paragraph*{Fatigue}  
Fatigue in MR helicopter training encompasses both musculoskeletal and visual strain arising from prolonged immersive exposure. Extended HMD use can increase neck and shoulder fatigue due to headset weight and repeated cockpit interactions, while sustained visual accommodation demands may contribute to eye strain and discomfort \citep{Souchet23a}. Visual fatigue is also associated with the VAC, where the visual system receives conflicting depth cues during immersive viewing \citep{Wang24}. Although studies report mixed findings regarding the extent to which VAC affects task performance \citep{Wang24,Mcanally24}, the literature consistently suggests that prolonged exposure to these perceptual conflicts may negatively influence comfort and training endurance. In helicopter pilot training, where long-duration sessions and continuous visual scanning are common, fatigue can reduce concentration, situational awareness, and overall training effectiveness. 

Several mitigation strategies have been proposed to manage fatigue while preserving simulation realism. Oculomotor exercises targeting vergence, saccades, and smooth pursuit may improve visual adaptation and reduce visual discomfort prior to immersive exposure \citep{Kim18,Park17}. These exercises demonstrate high contextual suitability and priority because they preserve simulator fidelity while supporting visual adaptation, although feasibility is rated medium due to the need for structured pre-exposure training procedures. High-quality HMD selection is also important, as advanced display systems and gaze-driven focal adjustment technologies may reduce visual fatigue associated with VAC, although feature-rich systems may simultaneously increase headset mass and musculoskeletal strain \citep{Yadin18,Wang21,Varjo,Cross22}. As discussed earlier, biometric monitoring may support fatigue/workload tracking, but should be treated as supplementary because physiological signals are subject to confounding factors. Limiting exposure duration may further reduce fatigue accumulation \citep{Cross22,Biwas24,Souchet23b}, although rigid session limits remain less compatible with the extended durations typical of helicopter simulator training. Overall, evidence supporting fatigue mitigation strategies is primarily derived from VR-based experimental and review literature with generally moderate RoB and limited MR-specific aviation validation. Nonetheless, the consistency of findings across ergonomic, physiological, and display-focused studies suggests these strategies are likely transferable to MR helicopter pilot training due to the shared visual and biomechanical demands of head-mounted immersive systems.

\paragraph*{Object angle location}  
The position and angle of virtual objects within immersive environments can contribute to musculoskeletal strain and fatigue, particularly when interaction requires repeated vertical head or neck movements \citep{Souchet23a}. Targets positioned above or below the natural line of sight may increase neck flexion or extension demands, contributing to discomfort during prolonged use. In MR helicopter pilot training, however, cockpit layouts, instrument positioning, and external visual references must closely replicate the real aircraft environment to preserve procedural fidelity and spatial realism. As a result, there is limited flexibility to reposition virtual objects for ergonomic optimization without compromising training authenticity. Accordingly, adjusting required movements or modifying object placement is rated high in feasibility but low in context suitability and priority. Although reducing repetitive head rotations or optimizing target placement may improve ergonomic comfort in general VR/MR applications \citep{Kim20b,Souchet23a,Botha24}, these adjustments are largely incompatible with aviation simulation requirements, where pilots must perform realistic cockpit scanning and operational viewing behaviors. Instead, ergonomic management is more appropriately addressed through complementary strategies such as improved posture guidance, HMD weight distribution, and cockpit setup optimization. The supporting evidence is primarily drawn from VR ergonomic studies and review literature with generally moderate RoB, but the underlying biomechanical principles are likely transferable to MR helicopter training due to the similar physical interaction demands of immersive HMD systems.

\section{Discussion}
\subsection{Summary of main findings}
MR HMDs have strong potential to increase immersion and reduce the cost burden associated with high-fidelity FSTDs, but the literature consistently shows that this promise is constrained by human-factor risks that can undermine training effectiveness. Across the included evidence base, cybersickness-related effects, visual fatigue, ergonomic strain, psychological and emotional adaptation factors, and broader sensory conflict emerged as the most recurrent barriers to sustained, operationally realistic use in pilot training contexts, thereby addressing RQ1. To address RQ2, eighteen human-factor drivers were synthesized and organized into a dual-taxonomy spanning individual, hardware, simulation-content, and ergonomic factors, supporting clearer reporting, comparison, and downstream design decision-making. Mitigation approaches identified in response to RQ3 spanned hardware, software, ergonomic, psychological, and physiological categories; however, their viability was strongly shaped by operational realism, fidelity requirements, and broader regulatory considerations. 

Several mitigation strategies emerged as the most actionable for training providers because they were rated high priority under our three-axis appraisal framework, directly addressing RQ4. Importantly, high priority did not require uniformly high feasibility and context-suitability scores; rather, strategies were prioritized when practical, operationally compatible, and targeted toward recurrent human-factor issues. Hardware-oriented strategies, particularly HMD choice, input type, and IPD calibration, consistently demonstrated strong feasibility and contextual suitability because they directly influence cybersickness, visual stability, ergonomic comfort, and perceptual alignment while preserving simulator fidelity. Similarly, ergonomic and psychological strategies, including postural instructions and monitoring, HMD weight management, onboarding with positive framing, and screening questionnaires, were prioritized because they can be integrated into existing instructor-led workflows with relatively low disruption to simulator realism or training procedures. In contrast, several content-modification strategies, including FOV restriction, peripheral blurring, reverse optical flow, and simulation intensity reduction, demonstrated effectiveness for reducing cybersickness but were generally rated lower in contextual suitability because they alter visual realism, peripheral awareness, or motion fidelity central to helicopter pilot training. 

Physiological strategies such as airflow, habituation, and oculomotor training appeared comparatively more operationally compatible because they support adaptation and comfort without substantially modifying simulator behavior. Biometric monitoring approaches, particularly eye tracking, pupillometry, and HRV monitoring, also showed strong operational promise due to their increasing validation as workload and discomfort indicators, alongside the growing availability of integrated and minimally intrusive sensing systems within modern HMD ecosystems. In contrast, more invasive or experimentally immature physiological approaches, such as GVS and pharmacological interventions, currently demonstrate lower operational maturity, greater implementation complexity, and more heterogeneous evidence bases. Overall, these findings suggest that mitigation suitability in MR helicopter pilot training depends not only on symptom-reduction effectiveness, but also on maintaining fidelity, workload realism, procedural authenticity, and compatibility with emerging regulatory expectations.

\subsection{Limitations}
Several limitations should be considered when interpreting this review. Although a systematic PRISMA-based process was followed, the synthesis may still underrepresent niche or emerging studies due to database selection, English-language restrictions, and the evolving terminology surrounding MR, VR, and AR systems. The evidence base also remains dominated by VR-derived studies, with comparatively fewer MR-specific aviation investigations, meaning that many conclusions rely on the transferability of shared HMD-mediated perceptual, visual, vestibular, and ergonomic mechanisms rather than direct MR helicopter-training validation. This transferability is supported by both theoretical and empirical literature suggesting that VR and MR systems share many underlying sensory-conflict pathways, particularly regarding latency, visual motion, optical alignment, workload, and ergonomic strain. However, MR environments additionally incorporate real-world visual references, passthrough systems, and hybrid interaction conditions that may alter the severity, manifestation, or operational significance of some human-factor effects. Accordingly, VR-derived findings should not be interpreted as perfectly equivalent to MR outcomes, but rather as the strongest currently available evidence base for informing emerging MR aviation applications. Further limitations relate to the methodological profile of the included studies. Across the evidence base, methodological rigor was variable, with recurring concerns relating to external validity, limited operational realism, heterogeneous evaluation methods, and inconsistent validation practices across empirical, technical, and review-level literature. Many studies were conducted in controlled laboratory or HCI-oriented environments rather than aviation-representative simulator settings, limiting confidence in direct transfer to regulated helicopter training pipelines. Consequently, the proposed taxonomy and mitigation ratings should be interpreted as structured evidence-informed guidance rather than definitive operational standards.

\subsection{Future directions}
A central implication of this review is that several mitigation strategies identified as most suitable for MR HMD helicopter pilot training now require direct validation within aviation-relevant simulator environments rather than continued inference from adjacent VR domains. Although the current evidence base supports a range of actionable interventions, the literature remains dominated by controlled quantitative studies and HCI/UX-oriented contexts, with comparatively limited longitudinal, qualitative, and operationally embedded aviation research. Consequently, future work should prioritize empirical validation studies in representative FSTD-style scenarios that evaluate high-priority mitigation bundles under realistic cockpit conditions, including instructor-led procedures, prolonged session durations, operational tasking, and fidelity constraints. Future studies should move beyond isolated single-factor demonstrations toward comparative and factorial experimental designs that benchmark against appropriate baselines, evaluate both comfort and training-relevant outcomes (e.g., cybersickness, workload, performance, retention, and situational awareness), and assess whether mitigation benefits persist across repeated sessions and different trainee populations. Particular attention should be given to strategies that appear operationally promising yet remain underexplored within aviation-specific MR contexts. Future work should also more explicitly examine where VR-derived findings remain transferable to MR and where inference has been made or important divergences emerge. 

Importantly, the findings synthesized throughout this review suggest that modern high-end MR HMDs increasingly approach the visual and perceptual capabilities of advanced VR systems, particularly regarding resolution, tracking stability, refresh performance, and gaze-driven optical features. Consequently, transferability findings between VR and MR human-factor outcomes may provide an evidence-based foundation for extending or adapting existing EASA-oriented VR HMD frameworks toward MR-specific special conditions, validation criteria, and fidelity requirements as the technology matures. Future regulatory development should therefore consider how current VR HMD special conditions may inform the creation of dedicated MR HMD qualification standards for helicopter pilot training.

\section{Conclusion}
MR HMDs present a promising pathway toward more immersive and potentially lower-cost helicopter pilot training; however, their integration remains constrained by significant human-factor challenges including cybersickness, visual fatigue, ergonomic strain, and sensory conflict. This review addressed four research questions relating to the identification of human-factor challenges, their underlying causes, mitigation strategies, and the operational suitability of those strategies within MR helicopter pilot training contexts. Eighteen human-factor drivers were synthesized into a dual-taxonomy linking underlying causes with corresponding mitigation strategies, providing a structured framework for understanding how different interventions interact with aviation-specific operational and regulatory constraints. The findings demonstrate that mitigation suitability in MR helicopter training depends not only on symptom reduction effectiveness, but also on compatibility with simulator fidelity, procedural realism, and workload demands. Strategies that preserved realism and integrated naturally into simulator workflows, such as high-quality HMD selection, airflow integration, physical cockpit controls, postural guidance, onboarding procedures, and habituation-based adaptation, were generally the most operationally suitable. 

In contrast, several cybersickness-reduction techniques that altered visual fidelity or motion perception showed lower contextual suitability despite experimental effectiveness. Although much of the current evidence base remains VR-derived, the shared perceptual, visual, vestibular, and ergonomic mechanisms across VR and MR HMD systems suggest meaningful transferability of many findings. As modern MR HMDs increasingly approach the perceptual capabilities of advanced VR systems, current VR-oriented aviation frameworks may provide an important foundation for future MR-specific qualification standards and operational guidance. Immediate next steps should focus on validating high-priority mitigation strategies within aviation-representative MR simulator environments, particularly through longitudinal and operationally embedded studies capable of informing future MR-specific regulatory standards. Overall, this review positions human-factor mitigation as essential for achieving safe, operationally credible, and regulatorily viable MR HMD helicopter pilot training systems.

\section*{Funding}
We thank Thales Australia for the financial support of this project through a top-up scholarship research grant and the Australian Research Council
(ARC) under Linkage grant LP240100414. 

\section*{Generative AI Declaration} Generative AI tools were used throughout preparation of this manuscript in a limited, copyeditor role. Specifically, OpenAI’s ChatGPT models (GPT-4o, o3, 5.2 and 5.5) were used to improve readability and language (e.g., grammar, wording, clarity, concision, consistency of terminology) and to assist with LaTeX formatting and layout. The tools were not used to generate content requiring original, creative, analytical, or critical thought. All AI-assisted edits were reviewed for accuracy and originality, and the authors remain accountable for the entire content of the manuscript.

\bibliographystyle{ArxivHarvard}
\bibliography{Bib2}


\end{document}